\documentclass[pre,aps,preprint,showpacs,byrevtex,superscriptaddress]{revtex4}
\usepackage{graphicx}

\begin{document}

\preprint{PREPRINT}

\title{Ion pairing in model electrolytes: A study via three particle correlation functions}
\author{Felipe Jim\'{e}nez-\'{A}ngeles}
\email{fangeles@www.imp.mx} \affiliation{Programa de
Ingenier\'{\i}a Molecular, Instituto Mexicano del Petr\'oleo,
L\'azaro C\'ardenas 152, 07730 M\'exico D. F., M\'exico}
\affiliation{Departamento de F\'{\i}sica, Universidad Aut\'onoma
Metropoloitana-Iztapalapa, Apartado Postal 55-334, 09340 M\'exico
D.F., M\'exico}
\author{Ren\'{e} Messina}
\email{messina@mpip-mainz.mpg.de} \affiliation{Institut f\"ur
Theoretische Physik II, Heirich-Heine-Universit\"at D\"usseldorf,
Universit\"atsstrasse 1, D-40225 D\"usseldorf, Germany}
\author{Christian Holm}
\email{holm@mpip-mainz.mpg.de}
\affiliation{Max-Planck-Institut f\"{u}r Polymerforschung, Ackermannweg 10, 55128 Mainz, Germany}
\author{Marcelo Lozada-Cassou}
\email{marcelo@www.imp.mx} \affiliation{Programa de
Ingenier\'{\i}a Molecular, Instituto Mexicano del Petr\'oleo,
L\'azaro C\'ardenas 152, 07730 M\'exico D. F., M\'exico}
\affiliation{Departamento de F\'{\i}sica, Universidad Aut\'onoma
Metropoloitana-Iztapalapa, Apartado Postal 55-334, 09340 M\'exico
D.F., M\'exico}

\date{\today{}}

\begin{abstract}
A novel integral equations approach is applied for studying ion
pairing in the restricted primitive model (RPM) electrolyte, i.
e., the three point extension (TPE) to the Ornstein-Zernike
integral equations.
In the TPE approach, the three-particle correlation functions
$g^{[3]}\left( {\bf r}_{1},{\bf r}_{2},{\bf r}_{3}\right)$ are
obtained. The TPE results are compared to molecular dynamics (MD)
simulations and other theories. Good agreement between TPE and MD
is observed for a wide range of parameters, particularly where
standard integral equations theories fail, i. e., low salt
concentration and high ionic valence. Our results support the
formation of ion pairs and aligned ion complexes.
\end{abstract}
\pacs{61.20.Qg}
\maketitle

\section{Introduction}

The restricted primitive model electrolyte (RPM) has been widely
studied by means of integral equations
\cite{rasaiah69,rasaiah70a,rasaiah70b,rasaiah72b,waismanL72a,waismanL72b,lozada83,croxton79}
and Monte Carlo (MC) simulations \cite{rasaiah72a,valleau_1970}.
All the approaches describe well the RPM in a wide regime of the
fluid phase diagram. Nonetheless, they {\em all fail} in the
dilute regime of a multivalent electrolyte
\cite{rasaiah72b,sjur83,lozada83}, which can be relevant for the
study of phase transitions in ionic fluids.

Such phase transition have been predicted as early as 1962, for
ionic mixtures \cite{mcquarrie1962}, and later for polyelectrolyte
solutions \cite{Belloni1986}. Experiments for the gas-liquid phase
transition of molten salts have been made in the past
\cite{kischembaum}. Among the first computer simulations of the
RPM, where this transition is reported are those of
Vorontsov-Vel'yaminov {\em et al}.
\cite{voronstov1970,voronstov1975}. Recently, there has been a
renewed interest in ionic phase transitions: Computer simulations
studies for the RPM
\cite{caillolJCP_1995,panaPRL_2002c,panaJCP_2002} and for
variations of this model, where unsymmetrical ionic charge and
size is considered
\cite{panaPRL_2000,depabloJCP_2001,depabloPRL_2001,panaPRL_2002b,panaPRE_2002a,depabloJCP_2002},
have given insight into the nature of the phase transition and the
molecular mechanisms behind these transitions. Experiments of a
liquid-liquid phase transitions have also been reported
\cite{friedman1950,sengersJPC_1990,pitzerJCP_1990,schroerJCP_1992,pitzerJPC_1994}.
Phase transitions of the RPM can be identified either with molten
salts gas-liquid transition or with the two liquid transition,
since in terms of dimensionless parameters the RPM does not
distinguish between these two sistems \cite{gonzalez1991}.

In the past, it has been proposed a powerful approach to
systematically incorporate correlations into any given liquid
theory
\cite{lozada81,lozada84,lozada90a,lozada90b,sanchez92,henderson92a}.
This method is known as {\em three point extension} (TPE) to
integral equations. By construction, TPE explicitly provides
valuable information of the three-particle correlations. In
consequence, the resulting pair distribution function includes
virtually infinitely more correlations and, hence, a better system
description is expected.
Although in the past TPE has not been applied to bulk fluids, our
presumption is sustained by previous TPE calculation for
inhomogeneous fluids where better agreement with computer
simulations \cite{alejandreMD,valleau91,lozadaPRE_1996} were
reported than in the case of {\em standard} \cite{note_IET}
integral equation theories \cite{lozada90a,lozada90b}.

In this paper we apply TPE to the RPM and compare with our
molecular dynamics (MD) simulations. Based on {\em ion pairing
association}, phase transitions in ionic fluids have been reported
by computer simulations \cite{panaPRE_2002a}. The results
presented here, (both TPE and our MD) for this region of the phase
diagram, support this ion paring association mechanisms. Moreover,
TPE, based on the agreement with our MD results, provides a
reliable theory to study ionic fluids in the important phase
diagram region of low ionic concentration and high coulombic
coupling.

In spite of important theoretical efforts made in the past, a
proper description of the full RPM electrolyte phase diagram is
still required. Previous approaches to study triplet correlations
have been developed by Kjellander {\em et al.} \cite{Kjellander-I}
and Plischke and Henderson \cite{Henderson-I,Henderson-II}. In
their study, they considered a fluid next to a plate and they
computed the inhomogeneous two-particle distribution function.
More relevant for the present study, however, is that of Attard
\cite{Attard-I1989,Attard-II1989}, who calculated the two particle
inhomogeneous distribution function (using the Percus-Yevick
closure) for a hard sphere fluid next to a hard sphere particle.
In his approach, he finds an excellent agreement with MC data. To
the best of our knowledge, no triplet correlation function has
been explicitly calculated for the RPM electrolyte.

In a study of the critical behavior of the RPM electrolyte at the
level of the Debye-H\"uckel theory, Levin and Fisher
\cite{fisher94,fisher96} have included triplet correlations by
imposing the presence of ionic pairs, such a consideration reveals
an Ising critical behavior. The ions pairs idea first proposed by
Bjerrum \cite{bjerrum} has been considerably extended by Levin and
Fisher, Stell and co-workers
\cite{stell_1972,stell1976,stellJCP_1995,stellJPC_1996,stellJP_2000}
and Blum and co-workers
\cite{blumJSP_1995,blumJCP_1996,blumMP_2001,blumJCP_2002}. While
ion pairing clearly seems to be the molecular mechanism ruling the
ionic solutions phase transitions
\cite{Suh1990,pitzerJCP_1990,depabloJCP_2001,panaPRE_2002a,panaPRL_2002c},
its physical bases remain unexplained \cite{panaPRE_2002a}. On the
other hand, although in the past some experiments supported a
classical critical behavior \cite{pitzerJCP_1990,sengersJCP_1992},
others the Ising universality class \cite{schroerJCP_1998} and
Singh and Pitzer \cite{pitzerJACS_1988} suggested a crossover from
classical to Ising behavior, later experimental results, however,
seem to agree in a crossover behavior
\cite{schroerJCP_1992,pitzerJPC_1994,pitzerPRL_1994,pitzerJCP_1995,sengersJCP_2001}.
Therefore, while ionic fluids asymptotic critical behavior appear
to exhibit ultimately Ising-like critical behavior, the question
of why do some ionic fluids appear to display classical behavior
\cite{fisher94}, remains unanswered \cite{sengersJCP_2001}. A
shortcoming of the ion pairing theories is that ion pairing is
imposed and hence they provide an {\em ad hoc} molecular
mechanism. Perhaps a better understanding of the molecular
mechanisms behind phase transitions could be captured by a formal
many body theory, such as TPE, where three particle correlations
are calculated explicitly, and no ion pairing is imposed.

In this work, by using the TPE to integral equations approach, we
obtain a better description of the RPM electrolyte: In particular
for the strongly coupled region. We also analyze the formation of
ion complexes.
The structure of the article is set out as follows. In
Sec.~\ref{sec.Theory} we present the TPE formalism. Section
\ref{sec.MD} is devoted to the computational details of the MD
simulation. In Sec.~\ref {sec.Results}, we present our results for
two typical (divalent) electrolyte concentrations. The obtained
three particle distribution function, $g_{\beta \gamma i
}^{[3]}\left( \mathbf{r}_{1},\mathbf{r}_{2},
\mathbf{r}_{3}\right)$, with TPE and MD simulation are compared
and analyzed in terms of ion asociation. We also compare the mean
force between two particles obtained with TPE, conventional
HNC/MSA and MD. Finally, Sec. \ref{sec.Conclusions} contains
concluding remarks.

\section{Theory}

\label{sec.Theory}

\subsection{Three point extension to integral equation theories}

\label{sec.Theory-TPE}

\begin{figure}[]
\includegraphics[width=12.0cm]{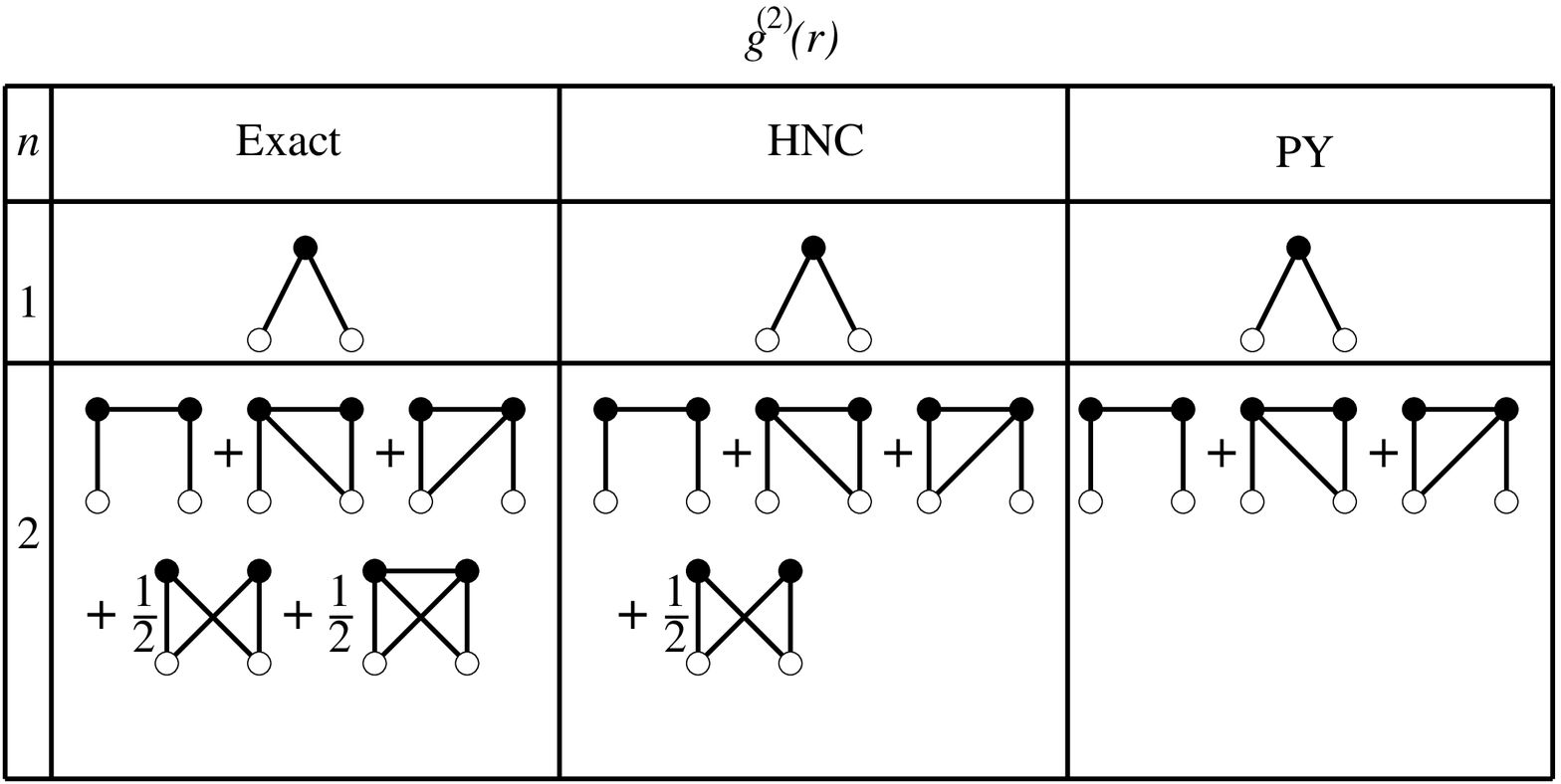}
\caption{Mayer diagrams for the first ($n=1$) and second ($n=2$) order in $%
\rho $ expansion of the pair correlation function, $g^{(2)}(r)$.
The exact, hypernetted chain (HNC) and Percus-Yevick (PY)
coefficients are shown. The black points and white dots are called
field and root points, respectively. The bonds represent the Mayer
function $f(\mathbf{r}_{12})$.} \label{mayer_Exp}
\end{figure}

The pair correlation function, $g(\mathbf{r}_{12}\equiv
\mathbf{r}_{1} - \mathbf{r}_{2})$, of a one-component fluid with
its components interacting through the pair potential
$u(\mathbf{r}_{12})$, is related to the potential
of mean force $w(\mathbf{r}_{12})$ (between two particles located at $%
\mathbf{r}_1$ and $\mathbf{r}_2$) by

\begin{equation}
g\left( \mathbf{r}_{12}\right) =\exp \left\{ -\beta w(\mathbf{r}%
_{12})\right\} .  \label{g2_w2}
\end{equation}
If $g\left( \mathbf{r}_{12}\right) $ is expanded in powers of the
bulk concentration, the $n$-th order coefficient is a sum of
integrals of
products of the Mayer function $f(\mathbf{r}_{12})\equiv \exp \{-\beta u(%
\mathbf{r}_{12})\}-1$. Such an integral of a product of Mayer
functions can be conveniently represented by Mayer diagrams \cite
{mcquarrie,friedman,henderson92a}. The diagrams of the first and
second order coefficients are given in the left hand side of
Fig.~\ref{mayer_Exp}. There is still not an exact theory to
compute $g(\mathbf{r}_{12})$, and all the available theories
ignore several classes of topologically different diagrams. We
will come back to this point below when we discuss the direct
correlation function and the Ornstein-Zernike equation.

In a multi-component fluid, the total correlation function,
$h_{ij}\left( \mathbf{r}_{12}\right) \equiv g_{ij}\left(
\mathbf{r}_{12}\right) -1$, between two particles of species $i$
and $j$ located at $\mathbf{r}_{1}$ and $\mathbf{r}_{2}$,
respectively, is related to the direct correlation function,
$c_{ij}\left( \mathbf{r}_{12}\right) $, through the
Ornstein-Zernike equation which for a $k$-component fluid is given
by

\begin{equation}
h_{ij}\left( \mathbf{r}_{12}\right) =c_{ij}\left(
\mathbf{r}_{12}\right) +\sum_{m=1}^{k}\rho _{m}\int h_{im}\left(
\mathbf{r}_{23}\right) c_{mj}\left( \mathbf{r}_{13}\right)
d\mathbf{r}_{3},  \label{oz1}
\end{equation}
where $\rho _{m}$ is the concentration of species $m$. Several
closures between $h_{ij}\left( \mathbf{r}_{12}\right)$ and
$c_{ij}\left( \mathbf{r} _{12}\right)$ have been proposed. For
instance,

\begin{equation}
c_{ij}\left(\mathbf{r}_{12}\right) =-\beta u_{ij}\left( \mathbf{r}
_{12}\right) +h_{ij}\left( \mathbf{r}_{12}\right) - \ln g_{ij}\left( \mathbf{%
r}_{12}\right),  \label{hnc1}
\end{equation}

\begin{equation}
c_{ij}\left( \mathbf{r}_{12}\right) =-\beta u_{ij}\left( \mathbf{r}%
_{12}\right) ,\quad \mathrm{and}  \label{msa1}
\end{equation}

\begin{equation}
c_{ij}\left( \mathbf{r}_{12}\right) =f_{ij}\left(
\mathbf{r}_{12}\right)
g_{ij}\left( \mathbf{r}_{12}\right) \exp \left\{ \beta u_{ij}\left( \mathbf{r%
}_{12}\right) \right\}.  \label{py1}
\end{equation}
Equations~(\ref{hnc1}), (\ref{msa1}) and (\ref{py1}) are known as
the hypernetted chain equation (HNC), the mean spherical
approximation (MSA) and the Percus-Yevick \ (PY) equation,
respectively. In the hypernetted chain theory, the \emph{bridge}
diagrams are ignored whereas in the Percus-Yevick approximation
both the bridge and \emph{product} diagrams are neglected \cite
{mcquarrie,hansen}. The first and second order Mayer graphs of the
HNC and PY theories are also given in Fig.~\ref{mayer_Exp}.

\begin{figure}[!h]
\includegraphics[width=12.0cm]{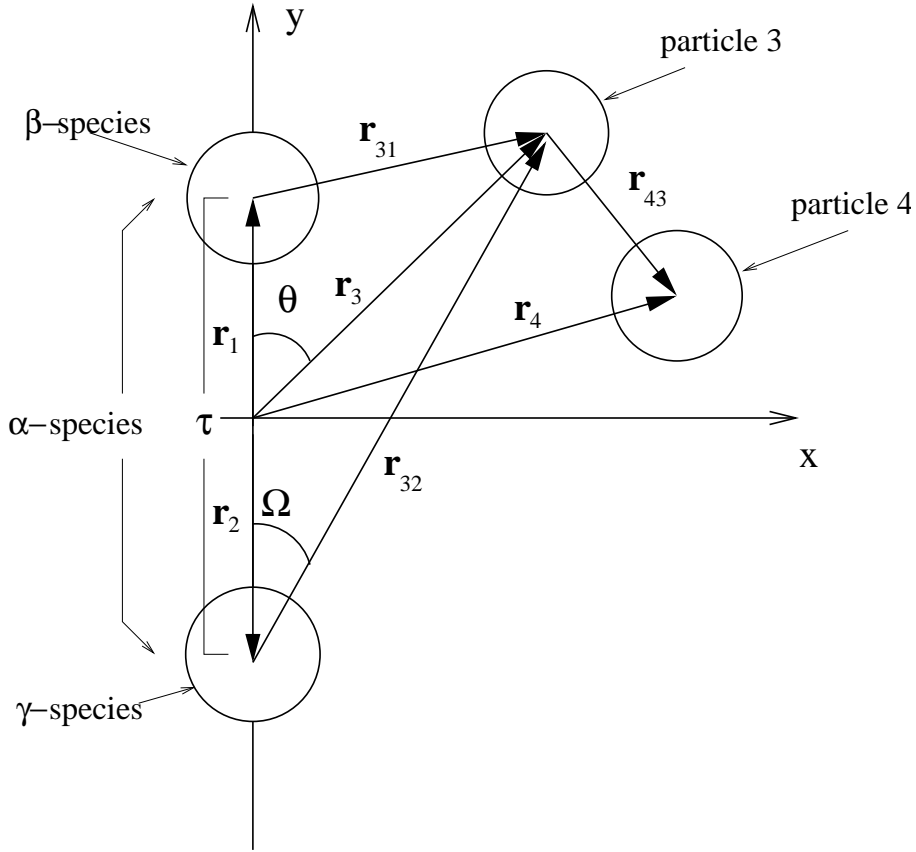}
\caption{Schematic representation of the model.} \label{setup}
\end{figure}

Let us now propose \cite{henderson92a,lozada84} that in a fluid of
$k$ -species there is an additional \emph{dumbbell} species at
infinite dilution made up of two particles (of species $\beta$ and
$\gamma$) at \emph{fixed} relative position
$\mathbf{t}\equiv\mathbf{r}_{12}$ (see Fig.~\ref{setup}). By
defining the dumbbell species as $\alpha$, we now have a $(k+1)$
-component fluid. For $\rho_{\alpha }\rightarrow 0$, the total
correlation function between the particle of species $\alpha$ and
the fluid particle of species $j$ reads

\begin{equation}
h_{\alpha j}\left( \mathbf{r}_{3}\right) =c_{\alpha j}\left( \mathbf{r}%
_{3}\right) +\sum_{m=1}^{k}\rho _{m}\int h_{\alpha m}\left( \mathbf{r}%
_{4}\right) c_{mj}\left( \mathbf{r}_{34}\right) d\mathbf{r}_{4},
\label{oz-tpe}
\end{equation}
where $c_{mj}\left( \mathbf{r}_{34}\right) $ is the direct
correlation
function between particles of species $m$ and $j$ both different from $%
\alpha $. In order to obtain $c_{mj}(\mathbf{r}_{34})$, the
$k$-component Ornstein-Zernike equation [Eq.~(\ref{oz1})] has to
be used. Different integral equation theories \cite{henderson92a}
can be obtained depending on the closure relations used for
$c_{\alpha j}\left( \mathbf{r}_{3}\right) $ and $c_{mj}\left(
\mathbf{r}_{34}\right) $ in Eq. (\ref{oz-tpe}). For
instance, TPE-HNC/MSA is obtained if MSA [Eq. (\ref{msa1})] is used for $%
c_{mj}\left( \mathbf{r}_{34}\right) $ and HNC [Eq.(\ref{hnc1})] for $%
c_{\alpha i}\left( \mathbf{r}_{3}\right) $.

In this formalism, the distribution function, $g_{\alpha i}\left( \mathbf{r}%
_{3}\right) $, of the $i$ species around the $\alpha $ species can
be interpreted as a conditional three-particle distribution
function denoted by $g_{\beta \gamma
i}^{[3]}(\mathbf{r}_{3};\mathbf{t}) \equiv g_{\beta \gamma
i}^{[3]}(\mathbf{r}_{1},\mathbf{r}_{2},\mathbf{r}_{3};\mathbf{t}=
\mathbf{r}_{1}-\mathbf{r}_{2})$, i.e., the density probability of
finding a particle of species $i$ at $\mathbf{r}_{3}$ in the
presence of the dumbbell.
Mathematically the conditional three particle distribution function, $%
g_{\beta \gamma i}^{[3]}(\mathbf{r}_{3};\mathbf{t})$, is related
to the
homogeneous three particle distribution function ${g}_{\beta \gamma i}^{(3)}(%
\mathbf{r}_{1},\mathbf{r}_{2},\mathbf{r}_{3})$ by

\begin{equation}
g_{\beta \gamma i}^{[3]}(\mathbf{r}_3;\mathbf{t})=
\frac{{g}_{\beta \gamma
i}^{(3)}(\mathbf{r}_1,\mathbf{r}_2,\mathbf{r}_3)} {g_{\beta \gamma}^{(2)}(%
\mathbf{t})}.  \label{g_three}
\end{equation}
The projection of ${g}_{\beta \gamma
i}^{[3]}(\mathbf{r}_1,\mathbf{r}_2, \mathbf{r}_3)$ gives directly
$g_{\beta \gamma }^{(2)}(\mathbf{t})$. This projection can be
provided by the Born-Green-Yvon theorem (BGY) that is based on a
balance of the mean effective force $F_{\beta \gamma}[g_{\beta
\gamma i}^{[3]}(\mathbf{r}_3;\mathbf{t})]$.

\subsection{The Born-Green-Yvon equation or a force law}

\label{sec.Theory-BGY}

\begin{figure}
\includegraphics[width=12.0cm]{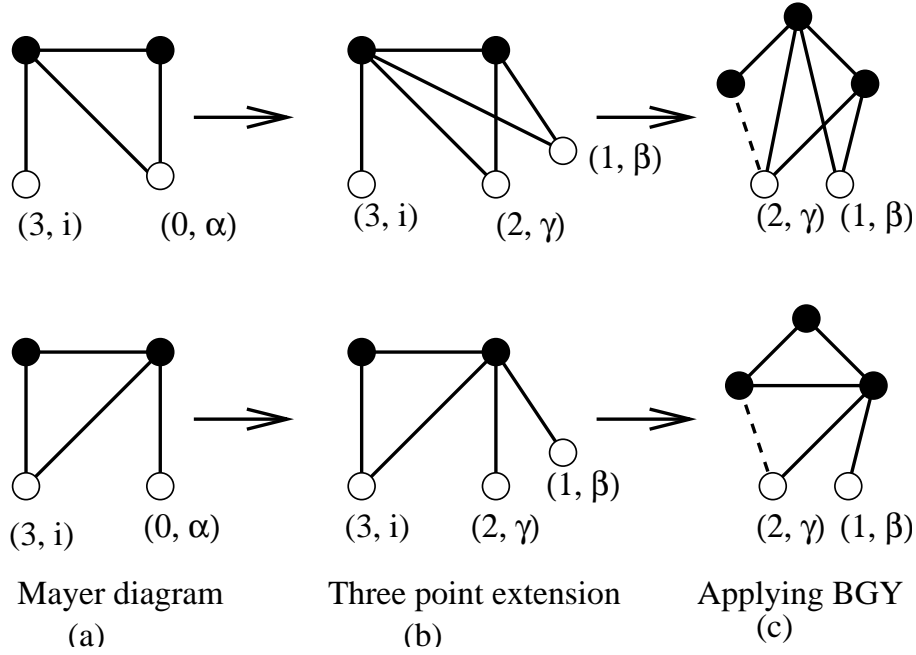}
\caption{Two examples of the transformation of Mayer diagrams
under TPE. The notation for the particles and species is the same
as in Fig.~\ref{setup}.
The dashed bond represents the $f_{\gamma i}(r_{23})=\frac{du_{\gamma i}(%
\mathbf{r}_{23})}{dr_{23}}$ function. Thereby, in ($N$,$\delta$),
$N$ stands
for the particle number and $\delta$ for the particle species (see also Fig.~%
\ref{setup}). $N=0$ stands for the dumbbell particle. (a) An
example of a second order Mayer diagram involving a dumbbell
particle and particle 3. (b) The same diagram as in (a) but the
constitutive particles of the dumbbell are explicited, i. e., at
the level of the triplet correlation function. (c) Resulting
diagrams upon applying BGY (on the level of the mean force).}
\label{Mayer_tpe}
\end{figure}

The Born-Green-Yvon (BGY) equation is one of the so called hierarchy
equations and it is an exact theorem relating the $n$ and $(n+1)$ particle
distribution functions \cite{born_green,yvon}. Here, we derive the BGY
equation as a sum of all the forces exerted on one of the two dumbbell's
particles (let us say particle of species $\gamma$ at $\mathbf{r}_2$). This
mean force has two contributions: (i) the direct force $\mathbf{f}_{\beta
\gamma }(\mathbf{t})$ exerted by the particle of species $\beta$ at $\mathbf{%
r}_{2}$ and (ii) the force $\mathbf{f}^{\mathrm{d}}_{\gamma }(\mathbf{t})$
exerted by all the other particles.
Thereby, the total mean force $\mathbf{F}_{\beta \gamma }(\mathbf{t})$ reads


\begin{equation}
\mathbf{F}_{\beta \gamma }(\mathbf{t})= \mathbf{f}_{\beta \gamma }(\mathbf{t}%
)+\mathbf{f}^{\mathrm{d}}_{\gamma }(\mathbf{t}),  \label{force1}
\end{equation}
Assuming that the dumbbell and fluid species are spherical particles
interacting through central force potentials, the component of $\mathbf{f}
_{\beta \gamma }$ along $\mathbf{t}$ is given by

\begin{equation}
\mathrm{f}_{\beta \gamma }(\tau\equiv|\mathbf{t}|)= -\frac{du_{\beta \gamma
}(\tau )}{d\tau},  \label{force2}
\end{equation}
where $u_{\beta \gamma}(\tau)$ is the potential of direct interaction
between the two dumbbell particles. The elementary force $d\mathbf{f}^{%
\mathrm{d}}_{\gamma}$ produced by a fluid element at $\mathbf{r}_3$ is given
by $d\mathbf{f}^{\mathrm{d}}_{\gamma}=\sum_{i=1}^{k}\mathbf{f}_{\gamma i} (
\mathbf{r}_{23}) \rho_{i}\left( \mathbf{r}_{3}\right) dv_3$, where $\mathbf{f%
}_{\gamma i}(\mathbf{r}_{23})$ is the force between a particle of species $i$
[of local density $\rho_i(\mathbf{r}_3)\equiv \rho_{i}g_{\beta \gamma
i}^{[3]}\left(\mathbf{r}_{3};{\tau}\right)$] at $\mathbf{r}_3$ and the
dumbbell's test particle of species $\gamma$. The component of $d\mathbf{f}^{%
\mathrm{d}}_{\gamma}$ along the direction of {$\mathbf{t}$} is given by

\begin{equation}
\begin{array}{l}
d\mathrm{f}^{\mathrm{d}}_{\gamma }  =  {\displaystyle \sum_{i=1}^{k}{\hat{%
\mathbf{t}}}\mathbf{\cdot f} _{\gamma i}\left(
\mathbf{r}_{23}\right) \rho _{i}\left( \mathbf{r}_{3}\right) dv_3
}
=  {\displaystyle -\sum_{i=1}^{k}\hat{\mathbf{t }}\mathbf{\cdot }\hat{%
\mathbf{r}}_{23} \frac{du_{\gamma i}(\mathbf{r}_{23})}{dr_{23}} \rho
_{i}\left( \mathbf{r}_{3}\right) dv_3,}
\end{array}
\label{force3}
\end{equation}
with {${\hat{\mathbf{t}}}$} and $\widehat{\mathbf{r}}_{23}$ being unit
vectors along the {$\mathbf{t}$} and $\mathbf{r}_{23}$ directions,
respectively, $u_{\gamma i}(\mathbf{r}_{23})$ is the potential of
interaction between an $i$-species particle with the $\gamma $-species
particle. 
Substituting Eqs.~(\ref{force2}) and (\ref{force3}) into Eq.(\ref{force1}), $%
\mathrm{F}_{\beta \gamma }$ is given by
\cite{lozada84,henderson92a}
\begin{equation}
\begin{array}{l}
\mathrm{F}_{\beta \gamma }(\tau) = {\displaystyle-\frac{dw_{\beta
\gamma }(\tau )}{d\tau } = - \frac{du_{\beta \gamma }(\tau)}{d\tau
}}   -  {\displaystyle \sum_{i=1}^{k}\rho _{i}\int
\frac{du_{\gamma i}(
\mathbf{r}_{23})}{dr_{23}}\cos \Omega g_{\beta \gamma i}^{[3]}\left(\mathbf{r%
}_{3}; {\tau}\right) dv_{3},}
\end{array}
\label{force4}
\end{equation}
where $\hat{\mathbf{t}}\mathbf{\cdot}\widehat{\mathbf{r}}_{23}=\cos\Omega $
and $w_{\beta \gamma}(\tau )$ is the potential of mean force between the two
dumbbell's particles. According to Eq. (\ref{g2_w2}),

\begin{equation}
w_{\beta \gamma}(\tau )= - k_{B}T \ln \left[ g_{\beta \gamma }(\tau) \right],
\label{mf_potential}
\end{equation}
and thus

\begin{equation}
\begin{array}{l}
{\displaystyle k_{B}T\frac{d\ln g_{\beta \gamma }(\tau)}{d\tau }}
{\displaystyle =-\frac{du_{\beta \gamma }(\tau)}{d\tau}}
{\displaystyle -\sum_{i=1}^{k}\rho _{i}\int \frac{du_{\gamma
i}(\mathbf{r} _{23})}{dr_{23}} \cos \Omega g_{\beta \gamma
i}^{[3]}\left( \mathbf{r}_{3};{\tau}\right) dv_{3},}
\end{array}
\label{bgy}
\end{equation}
which is the Born-Green-Yvon (BGY) equation. The degree of accuracy of $%
g_{\beta \gamma }(\tau )$ depends on the method used to compute
$g_{\beta \gamma i}^{[3]}\left( \mathbf{r}_{3};{\tau }\right)$. If
$g_{\beta \gamma i}^{[3]}\left( \mathbf{r}_{3};{\tau }\right)$ is
computed through the TPE of integral equation theories, it was
shown that new diagrams are included in the cluster expansion of
$g_{\beta \gamma}^{(2)}(\tau)$ \cite{henderson92a} (see
Fig.~\ref{Mayer_tpe}). By examination of the transformation of the
Mayer diagrams through the formalism outlined above, the
denomination of \emph{three point extension} becomes clear. A more
detailed description of TPE can be found in
ref.~\cite{henderson92a}.

\subsection{Application to the RPM electrolyte}

\label{sec.Theory-model} In the RPM electrolyte the fluid is considered as
made up of hard spheres of diameter $a$ with a central charge $q_i=z_{i}e$,
where $z_{i}$ is the valence of species $i$ and $e$ is the protonic charge.
The electroneutrality condition for the $n$-component electrolyte is

\begin{equation}
\sum_{i=1}^{n}z_{i}\rho_i =0.
\end{equation}

Assuming that the dumbbell particle ($\alpha$ species) is made up of two
particles of the same species from that in the fluid (see Fig. \ref{setup}),
the TPE-HNC/MSA equations become

\begin{equation}
g_{\alpha i}(\mathbf{r}_{3}) =\exp \left\{ -\beta u_{\alpha
i}(\mathbf{r}_{3})+\sum_{m=1}^{k}\rho_{m}\int h_{\alpha m}(
\mathbf{r}_{4}) c_{mi}(\mathbf{r}_{34}) d\mathbf{r}_{4}\right\},
\label{tpe-hnc-msa}
\end{equation}

where 
\begin{eqnarray}
u_{\alpha i}\left( \mathbf{r}_{3}\right)& = & u_{\alpha i}\left(
r_{13},r_{23}\right)  \nonumber \\
&= & \left\{
\begin{array}{l}
{\displaystyle \frac{z_{i}z_{\beta }e^{2}}{\varepsilon r_{23}}}+ {%
\displaystyle \frac{z_{i}z_{\gamma}e^{2}}{\varepsilon r_{13}}} \hspace{0.2cm}%
\mathrm{if} \hspace{0.2cm} r_{13} \hspace{0.2cm} \mathrm{and}\hspace{0.2cm}
r_{23}> a \\
\\
\infty \hspace{2.4cm} \mathrm{if } \hspace{0.2cm} r_{13} \hspace{0.2cm}
\mathrm{or } \hspace{0.2cm} r_{23}\leq a
\end{array}
\right.
\end{eqnarray}
with $z_{\beta }$ and $z_{\gamma }$ standing for valence number of particles
$\beta$ and $\gamma $, respectively. For spherical ions the direct
correlation function depends only of the ions distance $r_{34}=\left|
\mathbf{r}_{34}\right| .$ Within the mean spherical approximation, its
analytical expression is

\begin{equation}
c_{mi}(r_{34}) =c^{{hs}}(r_{34}) +z_{m}z_{i}c^{{sr}}(r_{34}) -\beta \frac{%
z_{m}z_{i}e^{2}}{\varepsilon r_{34}},
\end{equation}
where $c^{hs}(r_{34})$ is the direct correlation function for a hard spheres
fluid in the PY approximation and $c^{sr}(r_{34})$ is a short ranged
function. Because of the symmetry around the dumbbell axis, it is convenient
to use prolate coordinates ($\eta$, $\xi$, $\phi$) \cite{sanchez92,arfken}
defined as follows

\begin{equation}
\begin{array}{l}
x={\displaystyle{\frac{\tau}{2}}} \sqrt{\left( \eta ^{2}-1\right) \left(
1-\xi ^{2}\right) }\cos \phi , \\
y={\displaystyle \frac{\tau}{2}} \sqrt{\left( \eta ^{2}-1\right) \left(
1-\xi ^{2}\right) }\sin \phi , \\
z={\displaystyle \frac{\tau}{2}} \eta \xi ,
\end{array}
\end{equation}
and where the volume element is given by

\begin{equation}
dv=\frac{\tau ^{3}}{8}\left( \eta ^{2}-\xi ^{2}\right) d\phi d\xi d\eta.
\end{equation}
The relative distance $r_{34}$ is then given by

\begin{eqnarray}
r_{34}^{2} & = & \frac{\tau ^{2}}{4}\left\{ \left( \eta _{3}^{2}-1\right)
\left( 1-\xi _{3}^{2}\right) +\left( \eta _{4}^{2}-1\right) \left( 1-\xi
_{4}^{2}\right) +\left( \eta _{3}\xi _{3}-\eta _{4}\xi _{4}\right) ^{2}  \right. \nonumber \\
&-& \left. 2\sqrt{\left( \eta _{3}^{2}-1\right) \left( 1-\xi
_{3}^{2}\right) \left( \eta _{4}^{2}-1\right) \left( 1-\xi
_{4}^{2}\right) }\cos \phi_4
\right\}.  \nonumber \\
\end{eqnarray}
In prolate coordinates the potential of electrostatic interaction, $u^{%
\mathrm{el}}_{\alpha i}$, between the dumbbell and one fluid ion of species $%
i$ can be conveniently rewritten as

\begin{equation}
u^{\mathrm{el}}_{\alpha i}\left( \eta ,\xi \right) = {\displaystyle \frac{%
2e^{2}}{\tau \varepsilon }} \left( {\displaystyle \frac{z_{\beta}}{\eta -\xi
}} + {\displaystyle \frac{z_{\gamma }}{ \eta +\xi }} \right),
\end{equation}
and Eq. (\ref{tpe-hnc-msa}) as

\begin{eqnarray}
g_{\alpha i}\left(\eta_{3},\xi _{3}\right) & = & g_{\beta \gamma
i}^{[3]} \left( \eta_{3},\xi _{3};\tau\right) =  \exp \left\{ -
{\displaystyle \frac{2\beta e^{2}}{\tau \varepsilon }}
\left({\displaystyle \frac{z_{\beta }}{\eta _{3}-\xi _{3}}} +
{\displaystyle
\frac{z_{\gamma }}{\eta _{3}+\xi _{3}}} \right) \right.  \nonumber \\
&+& \int_{-1}^{1}\int_{\eta _{0}(\xi_4)}^{\infty }\rho _{\alpha
s}\left( \eta _{4},\xi _{4}\right)  \mathrm{{K}\left( \eta
_{3},\xi _{3},\eta _{4},\xi _{3}\right)
d\eta _{4}d\xi _{4}}  \nonumber \\
& +&z_{i}\int_{-1}^{1}\int_{\eta _{0}(\xi_4)}^{\infty }\rho _{\alpha \mathrm{%
{d} }}\left( \eta _{4},\xi _{4}\right)  \mathrm{{L}\left( \eta
_{3},\xi _{3},\eta_{4},\xi _{4}\right) d\eta
_{4}d\xi _{4}}  \nonumber \\
&-&\left. z_{i}\int_{-1}^{1}\int_{\eta _{0}(\xi_4)}^{\infty }\rho _{\alpha \mathrm{%
\ d}}\left( \eta _{4},\xi _{4}\right) \mathrm{{A}\left( \eta
_{3},\xi _{3},\eta _{4},\xi _{4}\right) d\eta _{4}d\xi _{4}}
-\mathrm{{J}\left( \eta _{3},\xi _{3}\right) }\right\},
\label{tpe-hnc-msa1}
\end{eqnarray}
with
\[
\eta_0(\xi)= \left\{
\begin{array}{ll}
\xi + b & \mathrm{for} \quad \xi_0 < \xi \le 1 \\
1 & \mathrm{for} \quad 0 \le \xi \le \xi_0
\end{array}
\right.
\]
and $\xi_0=1-b$, $b\equiv 2a/\tau$ and $\eta_0(-\xi)=\eta_0(\xi)$. The
expressions for K, L, A, J, $\rho_{\alpha \mathrm{s}}$ and $\rho_{\alpha
\mathrm{d}}$ are 
\begin{eqnarray}
\mathrm{K}( \eta _{3},\xi _{3},\eta _{4},\xi _{4}) &=&\frac{\tau ^{3}}{8}(
\eta _{4}^{2}-\xi _{4}^{2}) \int_{0}^{\phi _{\max }}c^{\mathrm{hs}}(r_{34})
d\phi_4 ,  \nonumber \\
\mathrm{L}( \eta_{3},\xi _{3},\eta _{4},\xi _{4}) &=&\frac{\tau ^{3}}{8}(
\eta_{4}^{2}-\xi _{4}^{2}) \int_{0}^{\phi _{\max }}c^{\mathrm{sr}}(r_{34})
d\phi_4 ,  \nonumber \\
\mathrm{A}( \eta_{3},\xi _{3},\eta _{4},\xi _{4}) &=&-\frac{\tau ^{4}\beta
e^{2}}{8\varepsilon }( \eta _{4}^{2}-\xi _{4}^{2}) \int_{0}^{2\pi }\frac{%
d\phi_4 }{r_{34}},  \nonumber
\end{eqnarray}
\[
\rho_{\alpha \mathrm{s}}\left(\eta_{4},\xi_{4}\right)= \rho_{\beta \gamma
\mathrm{s}}\left(\eta_{4},\xi _{4}\right) \equiv \sum_{m=1}^{n}\rho
_{m}h_{\alpha m}\left(\eta _{4},\xi _{4}\right),
\]
\[
\rho_{\alpha \mathrm{d}}\left(\eta_{4},\xi_{4}\right)= \rho_{\beta \gamma
\mathrm{d}}\left(\eta_{4},\xi _{4}\right) \equiv \sum_{m=1}^{n}z_{m}\rho
_{m}h_{\alpha m}\left(\eta _{4},\xi _{4}\right),
\]
\begin{eqnarray*}
\rm{J}\left( \eta _{3},\xi _{3}\right) = \int_{-1}^{-\xi _{\rm
min}(\tau)} \int_{1}^{\eta _{0}(\xi_4)}{\rm K}\left( \eta _{4},\xi
_{4},\eta _{3},\xi _{3}\right) d\eta _{4}d\xi_{4} +\int_{\xi _{\rm
min}(\tau) }^{1}\int_{1}^{\eta _{0}(\xi_4)}{\rm K}\left( \eta
_{4},\xi _{4},\eta _{3},\xi _{3}\right) d\eta _{4}d\xi _{4},
\end{eqnarray*}
respectively, with 
\[
\xi_{\mathrm{min}}(\tau)= \left \{
\begin{array}{ll}
0 & \mathrm{if}\quad \tau \le a \\
\xi_0 & \mathrm{if}\quad \tau > a.
\end{array}
\right.
\]
By introducing the elliptic function of second kind $\mathrm{F}(\pi/2,k)$,
one can rewrite A as
\begin{equation}
\mathrm{A}\left( \eta _{3},\xi _{3},\eta _{4},\xi _{4}\right) = {\ %
\displaystyle \frac{\tau \left( \eta_3 ^{2}-\xi_3 ^{2}\right) \mathrm{F}
\left( \pi /2,k\right)}{2r_{34}^{max}}}
\end{equation}
where
\begin{equation}
k^{2}=\frac{\tau^{2}\sqrt{\left( \eta_{3}^{2}-1\right) \left( 1-\xi
_{3}^{2}\right) \left( \eta _{4}^{2}-1\right) \left( 1-\xi _{4}^{2}\right) }
}{2(r^{max}_{34})^{2}},
\end{equation}
and
\begin{eqnarray}
\left( r^{max}_{34}\right)^2 & = & \frac{\tau^2}{4}\left[
\sqrt{\left( \eta_{3}^{2}-1\right) \left( 1-\xi _{3}^{2}\right)}
+\sqrt{\left( \eta_{4}^{2}-1\right) \left( 1-\xi _{4}^{2}\right) }
\right] ^{2} + \left( \eta_{3}\xi_{3}-\eta_{4}\xi_{4}\right) ^{2}.
\end{eqnarray}
Eq. \ref{tpe-hnc-msa1} is in fact a set of two coupled, three
dimensional, non-linear integral equations. To solve these
equations, we have developed a sophisticated, but efficient,
finite element method for its solution (see appendix for details
on our numerical method).

Using Eq. (\ref{force4}), the mean force between the two dumbbells
particles reads

\begin{equation}
F_{\beta \gamma }(\tau )=f_{\beta \gamma }^{\ast }\left( \tau \right)
+f_{\beta \gamma }^{\mathrm{el}}\left( \tau \right),  \label{net_force}
\end{equation}
where

\begin{eqnarray}
f_{\beta \gamma }^{\ast }\left( \tau \right) & = & \frac{\pi \tau
^{2}}{ 2\beta } \sum_{j=1}^{2}\rho
_{j}\int_{\xi_{\mathrm{min}}(\tau)}^{1}g^{[3]}_{\beta \gamma j}
\left[\eta_0\left( \xi_3 \right) ,\xi_3;\tau\right] \left[ -2\xi_3
^{3}-3b\xi_3 ^{2}+ \left(2-b^{2}\right) \xi_3 +b\right] d\xi_3
\label{contact_force}
\end{eqnarray}
and
\begin{eqnarray}
f_{\beta \gamma }^{\mathrm{el}}( \tau ) & = & {\displaystyle
\frac{z_{\beta }z_{\gamma_3}e^{2}}{\varepsilon \tau ^{2}}}+
{\displaystyle \frac{\tau \pi z_{\beta }e^{2}}{\varepsilon }}
\int_{-1}^{1}\int_{\eta_{0}\left( \xi_3 \right) }^{\infty
}\rho_{\beta \gamma \mathrm{d}} ( \eta_3 ,\xi_3 ) 
\frac{( 1-\eta_3 \xi_3 ) ( \eta_3 +\xi_3) }{( \eta_3 -\xi_3 )
^{2}} d\xi_3 d\eta_3.  \label{electrical_force}
\end{eqnarray}
Thus the pair distribution function of the electrolyte solution is given by

\begin{equation}
g_{\beta \gamma}(r) = \exp \left\{-\beta \int_{\infty }^{r}F_{\beta \gamma
}(\tau)d\tau \right \}.
\end{equation}

The solution of Eq.~(\ref{tpe-hnc-msa1}) and calculation of
$F_{\beta \gamma} $ through Eqs.~(\ref{net_force}),
(\ref{contact_force}) and (\ref {electrical_force}) were
numerically solved.

\section{Molecular dynamics}

\label{sec.MD}

The electrolyte is confined in a cubic box of length L. The bulk
salt concentration $\rho $ is then given by ${\displaystyle
\frac{N}{L^{3}}}$, where $N$ is the number of positive (or
negative) ions. The dumbbell is made up of two $fixed$ ions (with
a center-center separation $\tau$) disposed symmetrically along
the axis passing by the two centers of opposite faces.
A similar system setup was also used elsewhere to study two fixed
macroions \cite{messina_PRL_00,messina_EPL_00,messina_PRE_01}. We
use MD simulations to compute the motion of the mobile fluid ions
coupled to a heat bath acting through a weak stochastic force
$\mathbf{W}_{i} \mathbf{(}t)$ with a zero mean value. The equation
of motion of any mobile ion $i$ reads

\begin{equation}
m\frac{d^{2}\mathbf{r}_{i}}{dt^{2}}=-\nabla _{i}U-m\Gamma
\frac{d\mathbf{r} _{i}}{dt}+\mathbf{W}_{i}(t),  \label{MD-newton}
\end{equation}
where $m$ is the ion mass, $\Gamma$ is the friction coefficient
and $-\nabla _iU$ is the potential force having two contributions:
(i) the Coulomb interaction and (ii) the excluded volume
interaction. Friction and
stochastic force are linked by the dissipation-fluctuation theorem $%
\left\langle \mathbf{W}_{i}(t)\cdot \mathbf{W}_{j}(t^{\prime
})\right\rangle =6m\Gamma k_{B}T\delta _{ij}\delta \left(
t-t^{\prime }\right)$.

Excluded volume interactions are modeled by a pure repulsive Leonard-Jones ($%
LJ$) potential defined by

\begin{equation}
U_{LJ}(r)=\left\{
\begin{array}{ll}
4\epsilon_{LJ} \left[ \left( {\displaystyle \frac{a}{r}}\right)
^{12}-\left( {\displaystyle \frac{a}{r}}\right) ^{6}\right]
+\epsilon_{LJ}, & \mathrm{for}
\hspace{0.2cm} r<2^{1/6}a \\
0, & \mathrm{for } \hspace{0.3cm}r\geq 2^{1/6}a ,
\end{array}
\right.  \label{eq.MD-LJ}
\end{equation}
where $a$ is the ion diameter.

The electrostatic interaction between any pair $ij$, where $i$ and
$j$ denote either a dumbbell ion and/or a mobile fluid ion, reads

\begin{equation}
U_{el}(r)={\displaystyle \pm k_{B}T\ell _{B}\frac{z^{2}}{r}},
\label{eq.MD-coulomb}
\end{equation}
where +(-) applies to ions likely(oppositely) charged,
${\displaystyle \ell _{B}=\frac{e^{2}}{\varepsilon k_{B}T}}$ is
the Bjerrum length describing the electrostatic strength and $z$
is the salt valence ($z_{i}=z_{j}=z$). To link our system
parameters to experimental units we choose the $LJ$ energy
parameter $\epsilon _{LJ}=k_{B}T$ (where $T=298\mathrm{K}$) and
$a=4.25$ \AA . This leads then to the water Bjerrum length $\ell
_{B}=1.68a=7.14$ \AA . A macroscopic system was mimicked by
imposing periodic boundary conditions. The long range Coulomb
interaction was treated by using an optimized and efficient Ewald
summation variant, namely the particle-particle-particle-mesh
(P3M) method \cite{deserno98a}.

In order to limit the size effects, we choose $L$ sufficiently
large, typically 10 times (or more) the Debye-H\"uckel screening
length. The number of ions in the box is 500 for all cases
(concentrated and dilute solutions). It is important to mention
that the computation of $g^{[3]}_{\beta \gamma i}(r,\theta;\tau)$
is statistically extremely demanding and especially for small
$\theta$ angles, since the quantity of information varies like
$\sin (\theta )$. In this notation, the distance
$r\equiv|\mathbf{r}_3|$ and the angle $\theta\equiv\angle
{(\mathbf{r}_1, \mathbf{r}_3)}$ are always relative to the center
of the dumbbell (see Fig.~\ref{setup}). The fact that the
observable $g^{[3]}_{\beta \gamma i}(r,\theta;\tau)$ concerns only
an ``elementary solid angle'', it strongly reduces the available
information compared to that available for the \textit{pair}
correlation function, since in that latter case a full solid angle
$4\pi$ and many ion pairs are accessible. To overcome this
difficulty, we considered a sufficiently large angle range $\Delta
\theta $ (typically between $5-15^0$ depending on the
concentration $\rho$), so that the gathered informations contains
as less noise as possible. On the other hand, $\Delta \theta $
must not to be too large otherwise the resolution gets too small.
For each system under consideration, a compromise between these
two effects that had to be found.

Finally for the computation of the effective mean force between
two ions, we considered the same system but where no fixed
dumbbell is present. Thereby, we could compute the potential of
mean force, knowing the $g(r)$, and then get by derivation the
effective force.

\section{Results}

\label{sec.Results}

We have done calculations for the 1:1 and 2:2 electrolytes using
the TPE-HNC/MSA integral equation. {\em For the 1:1 electrolyte
the agreement between TPE-HNC/MSA and MD results is qualitative
and quantitatively very good}. However, in order to keep low the
number of plots we just present a detailed analysis on the results
of the 2:2 electrolyte. The choice of divalent ions is motivated
by the fact that it represent a strong test for liquid theories.
Thereby, we considered two typical concentrations: (i) the
concentrated case with $\rho=1$M and (ii) the dilute case with
$\rho=0.005$M. As a main result, the effective mean force obtained
by TPE-HNC/MSA, HNC/MSA, and MD simulation is presented for each
concentration regime. In order to further quantify the robustness
of the TPE-HNC/MSA theory, we investigated in detail the
conditional three-particle distribution function, $g_{\beta \gamma
i}^{[3]}(r,\theta;\tau )$, by comparing TPE-HNC/MSA with MD.

For the discussion, it is convenient to adopt the following
notations: $g_{++-}^{[3]}(r,\theta ;\tau )$ stands for the
distribution function of negative ions when the dumbbell is made
up of two positive ions, $g_{+--}^{[3]}(r,\theta ;\tau )$ for that
of negative ions when the dumbbell is made up of a negative and a
positive ions, and so on. By symmetry the three particle
distribution function satisfies $g_{++-}^{[3]}(r,\theta ;\tau
)=g_{--+}^{[3]}(r,\theta ;\tau )$ and also $g_{+--}^{[3]}(r,\theta
;\tau )=g_{+-+}^{[3]}(r,\pi -\theta ;\tau )$. Thereby, we
systematically compared theory and simulation for $g_{\beta \gamma
i}^{[3]}(r,\theta ;\tau)$, but show results only for $g_{\beta
\gamma i}^{[3]}(r,\theta ;\tau =a)$ (i.e., when the two dumbbell
ions are in contact), for two given values of $\theta $ ($\pi /4$
and $\pi /2$). In addition, within the TPE-HNC/MSA theory, we also
provide the full $\theta $-dependence of $g_{\beta \gamma
i}^{[3]}(r,\theta ;\tau )$ for different $\tau $.

\subsection{Concentrated case}

\label{sec.Results-concentrated}

In this section, we deal with the concentrated electrolyte
solution ($\rho=1$ M). The electrostatic screening at such high
ionic density and valence ($z=2$ ) is very strong. The study of
such a system is important to test TPE-HNC/MSA theory, since
already inhomogeneous and homogeneous HNC/MSA theories are in
excellent agreement with molecular simulations under such
conditions \cite{degreve93}.

\subsubsection{Three particle correlation function}

\paragraph{Symmetric dumbbell}

\begin{figure}
\includegraphics[width=12.0cm]{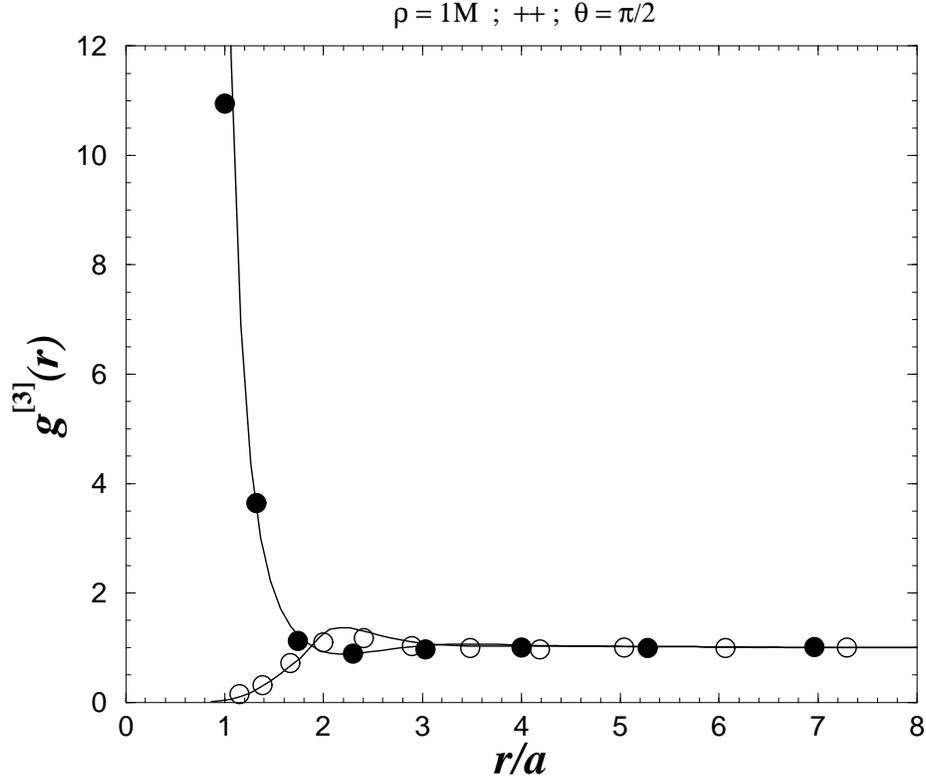}
\caption{Three particle distribution function $g_{++i}^{[3]}(r,\theta =\pi %
/2;\tau =a)$ for a dumbbell made of two (divalent) positive particles, with $%
\rho =1M$ and $z=2$. The solid lines represent the results from
TPE-HNC/MSA.
The MD results are shown in filled and open circles for $g_{++-}^{[3]}(r,\pi %
/2;a)$ and $g_{+++}^{[3]}(r,\pi /2;a)$ respectively.}
\label{fig.1M_90_++}
\end{figure}

We first consider symmetric dumbbells made of like charged
positive divalent
ions. The profiles of $g_{++-}^{[3]}(r,\theta =\pi /2;\tau =a)$ and $%
g_{+++}^{[3]}(r,\theta =\pi /2;\tau =a)$ are depicted in Fig. \ref
{fig.1M_90_++}. Concerning the negatively charged fluid ions
(i.e., ``dumbbell counter-ions''), we have quantitative agreement
between theory and simulation even near the distance of closest
approach. The slight difference at short distance ($r\sim a$) is
due to the fact that for the short-ranged excluded volume
interaction, MD simulation is built with a soft-core $LJ$
potential whereas the actual theory uses a true hard-core
potential. For the positively charged fluid species (``dumbbell
co-ions''), we also have an excellent qualitative agreement. The
TPE-HNC/MSA maximum of the co-ion distribution function is within
the statistical error, however slightly higher than MD data. The
location of the maximum is nearly the same as that found with
simulation. Hence TPE-HNC/MSA has an excellent agreement with MD,
within the numerical error.

\begin{figure}
\includegraphics[width=12.0cm]{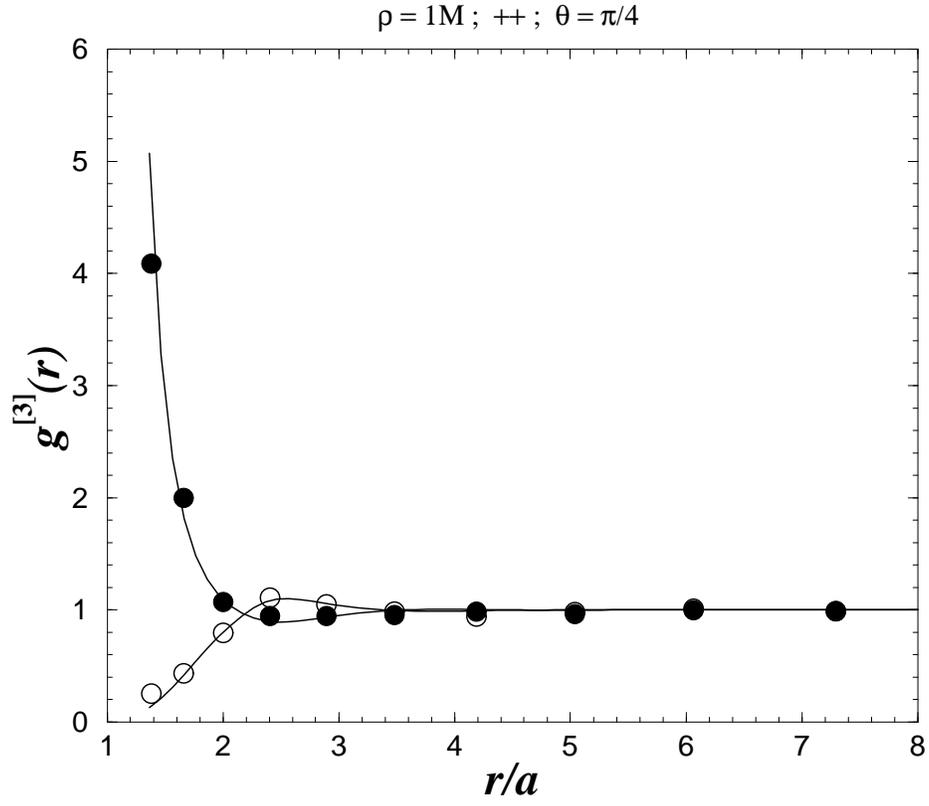}
\caption{Same as in Fig.\ref{fig.1M_90_++} with $\theta = \pi/4$.}
\label{fig.1M_45_++}
\end{figure}

For $\theta =\pi /4$ (see Fig. \ref{fig.1M_45_++}), one still has
the same quantitative agreement between MD and TPE-HNC/MSA for the
dumbbell counter-ions. It is observed that the value of
$g_{++-}^{[3]}(r,\pi /4;a)$ at closest approach ($r=1.29a$) is not
as high as at $\theta =\pi /2$ (see Fig. \ref {fig.1M_90_++}) for
the corresponding plot. The physical reason of this feature is
straightforward. The closest approach to the center of the
dumbbell is larger at $\theta =\pi /4$ than at $\theta =\pi /2$,
therefore, since all particles have the same size, the resulting
$attractive$ electrostatic interaction between the dumbbell and
the counter-ion is higher at $\theta =\pi /2$. For the
dumbbell-co-ions distribution $g_{+++}^{[3]}(r, \pi /4;a)$ we have
quantitative agreement between TPE-HNC/MSA and MD.

\begin{figure}
\includegraphics[width=12.0cm]{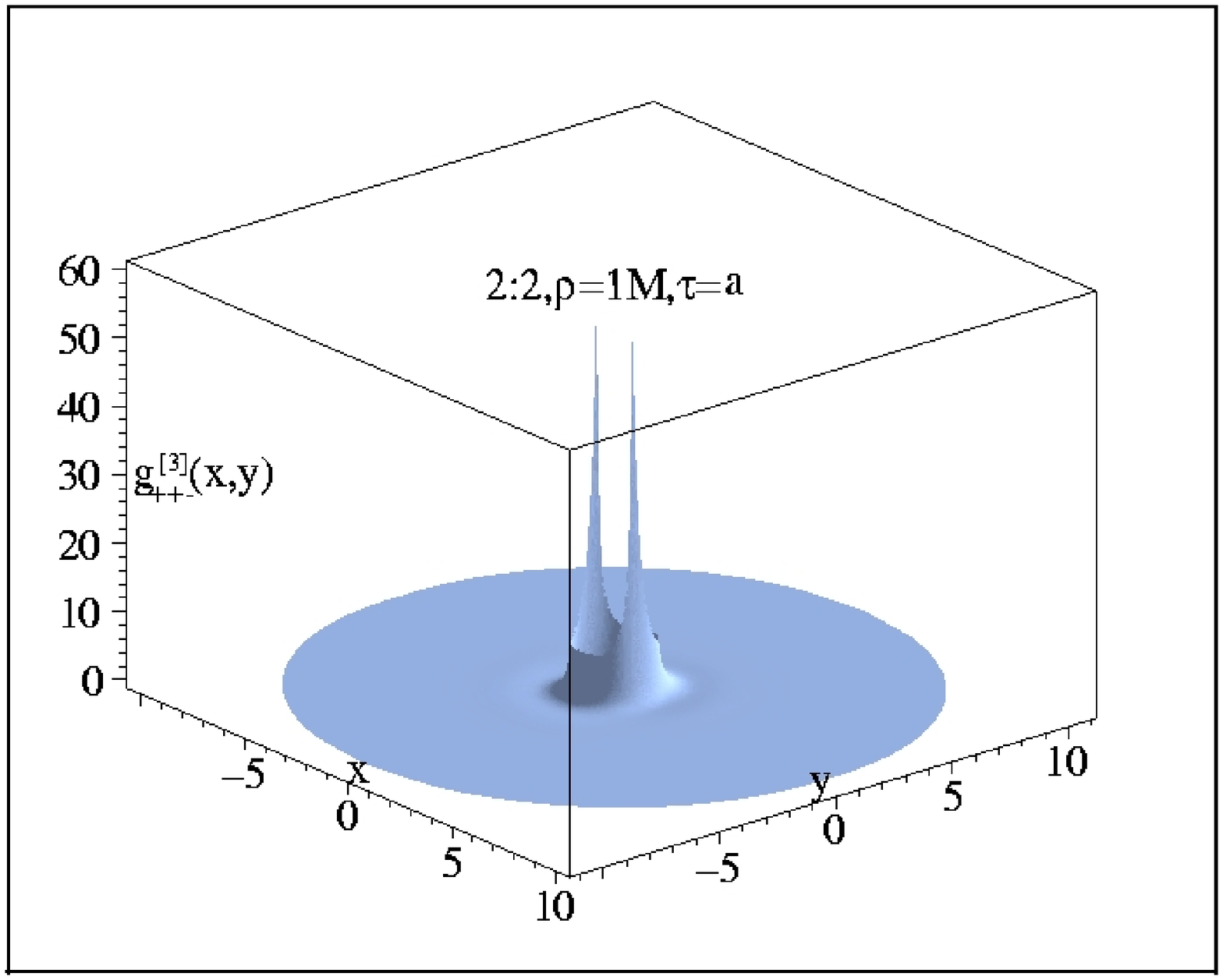} %
\includegraphics[width=12.0cm]{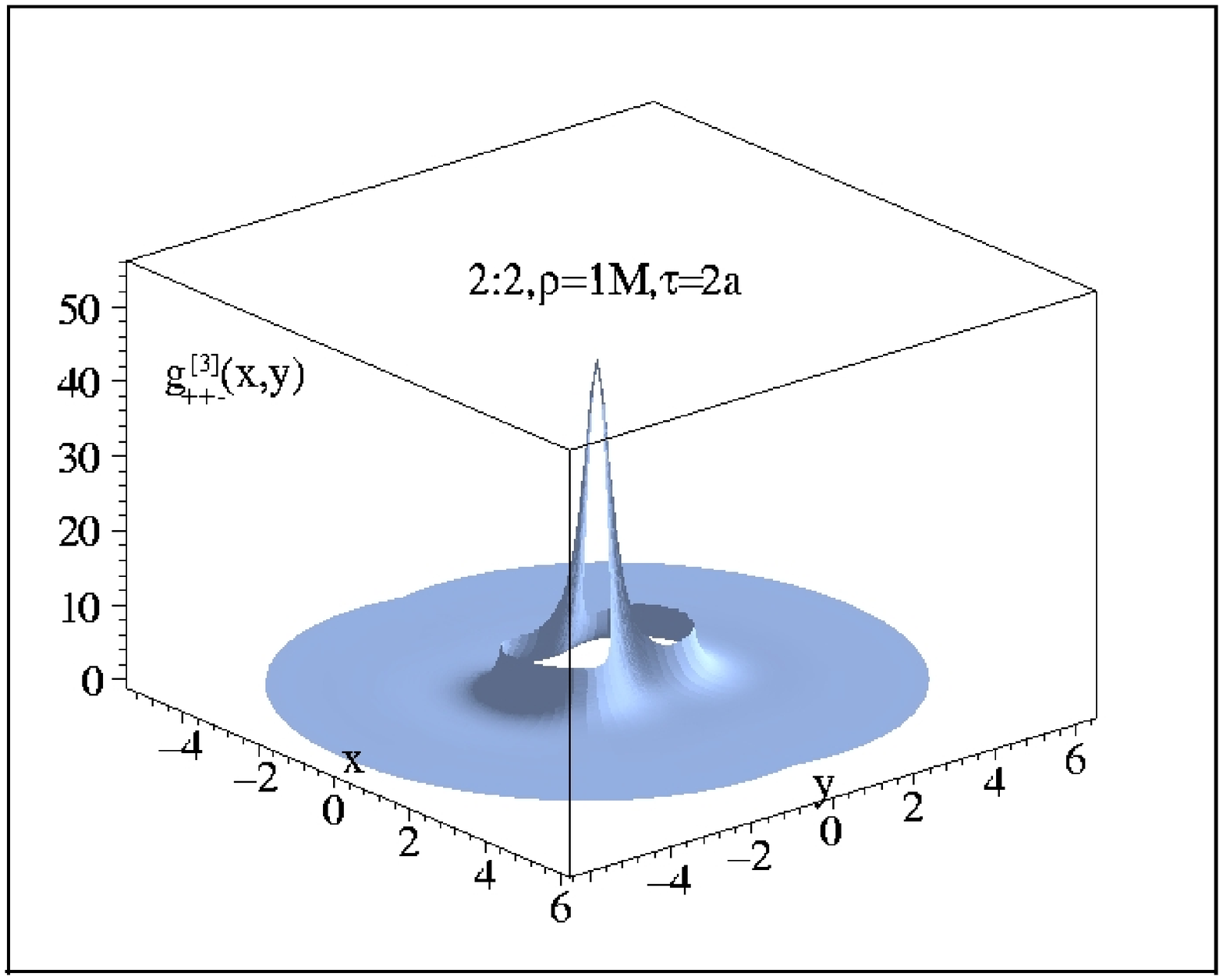}
\caption{3D representation (in Cartesian coordinates) of: (a)
(upper 3D
plot) $g_{++-}^{[3]}(r,\theta ;\tau =a)$ and (b) (lower 3D plot) $%
g_{++-}^{[3]}(r,\theta ;\tau =2a)$ obtained by TPE-HNC/MSA for the
same fluid parameters as in Figs.~\ref{fig.1M_90_++} and
\ref{fig.1M_45_++}. The dumbbell axis is parallel to $y$ axis.}
\label{fig.plot3D_1M_++}
\end{figure}

The three dimensional (3D) plots of the three particle
(counter-ion-dumbbell) distribution function
$g_{++-}^{[3]}(r,\theta ;\tau )$
obtained by TPE-HNC/MSA are sketched in Fig. \ref{fig.plot3D_1M_++}. At $%
\tau =a$ (dumbbell ions at contact), Fig.
\ref{fig.plot3D_1M_++}(a) shows a strong variation near to the
surface of closest approach. As expected, the maximum is obtained
at $\theta =\pi /2,3\pi /2$ ($g_{++-}^{[3]}\approx 50$), whereas
the minimum is at $\theta =0,\pi $ ($g_{++-}^{[3]}\approx 8$).
Moreover, we have oscillations in the distribution function, as a
function of $r$, for any $\theta$ , which confirms our previous
observations at $\theta =\pi /2,\pi /4$ (see Figs.
\ref{fig.1M_90_++} and \ref{fig.1M_45_++}). We have carefully
checked that this feature holds at any $\tau $. The 3D plot of the
co-ion-dumbbell distribution $g_{+++}^{[3]}(r,\theta ;a)$ is not
reported here.

At a larger dumbbell separation, $\tau =2a$ [see Fig.
\ref{fig.plot3D_1M_++} (b)], $g_{+++}^{[3]}(r,\theta ;2a)$ is
still highly peak at $\theta =\pi /2$ and has its maximal value at
the middle point of the dumbbell. At sufficiently large
separation, we have an isotropic counter-ion distribution around
each dumbbell particle (not shown here). We point out
that although the probability of finding two like-charged ions in contact ( $%
\tau =a$) is very low, the probability of having more than two
counter-ions in contact (at $\theta \approx \pi /2$) with the two
like-charged ions dumbbell, is very high. This implies an
overcompensation of the dumbbell's charge, which is verified by
the observed oscillations in the counter-ions profile of Fig.\ref
{fig.plot3D_1M_++}, since oscillations imply an electrical field
inversion, which implies charge reversal. It should be stressed
that to calculate thermodynamics functions such as the internal
energy or pressure, $g^{[3]}_{++i}(r,\theta;\tau)$ for every
$\tau$ must be known, even if the probability of finding two
like-charged ions at contact is very low.

\paragraph{Antisymmetric dumbbell}

\begin{figure}
\includegraphics[width=12.0cm]{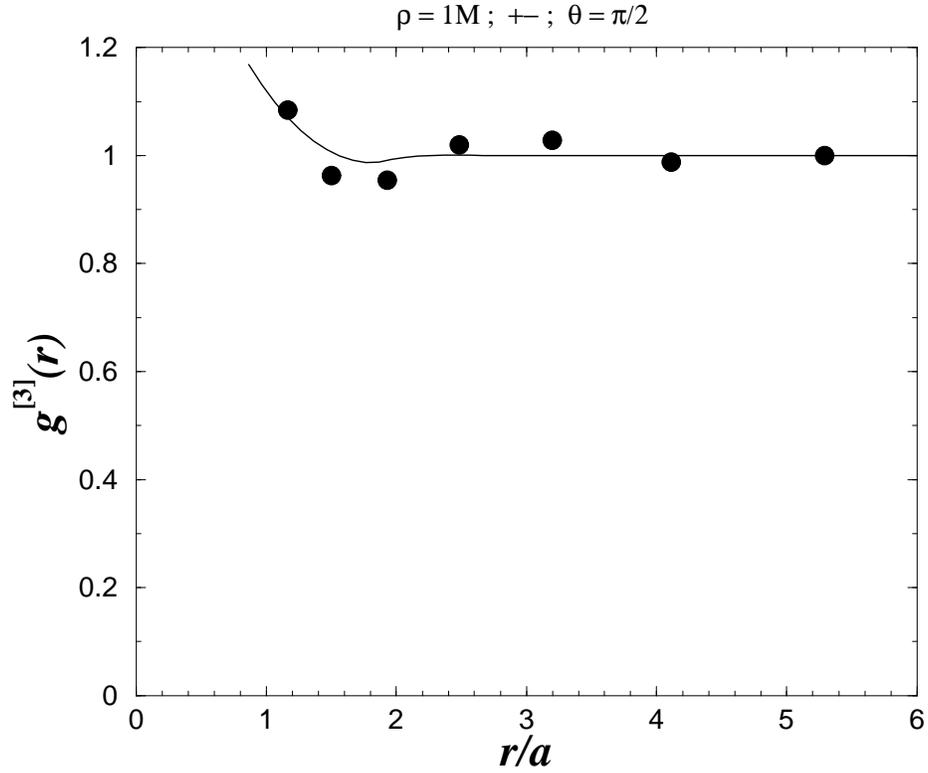}
\caption{Three particle distribution function
$g_{+-i}^{[3]}(r,\theta =\pi/2;\tau =a)$ for a dumbbell made of a
positive and a negative divalent ions, with $\rho =1M$ and $z=2$.
The solid lines represent the results from
TPE-HNC/MSA. The MD results are shown in filled circles. The curves for $%
g_{+--}^{[3]}(r,\pi /2;a)$ and $g_{+-+}^{[3]}(r,\pi /2;a)$ colapse
in a single curve.} \label{fig.1M_90_+-}
\end{figure}

We now consider antisymmetric dumbbells made of two opposite
divalent ions.
In this case, by symmetry arguments we expect that $g_{+--}^{[3]}(r,\pi %
/2;a)=g_{+-+}^{[3]}(r,\pi /2;a)$. The profiles of
$g_{+--}^{[3]}(r,\pi /2;a)$ and $g_{+-+}^{[3]}(r,\pi /2;a)$ are
plotted in Fig. \ref{fig.1M_90_+-}. Since at $\theta =\pi /2$ the
electric field component (produced by the dumbbell) perpendicular
to the dumbbell axis is zero, the electrostatic correlations are
only generated by the fluid ions. Consequently, we expect a
quasi-neutral fluid behavior. This is precisely what Fig. \ref
{fig.1M_90_+-} shows for theory and simulation, where $g_{+--}^{[3]}(r,\pi %
/2;a)$ and $g_{+-+}^{[3]}(r,\pi /2;a)$ curves collapse in a single
curve. The adsorption at contact, is a hard sphere entropic effect
due to the salt high concentration. This adsorption does not occur
at low salt concentration.

\begin{figure}
\includegraphics[width=12.0cm]{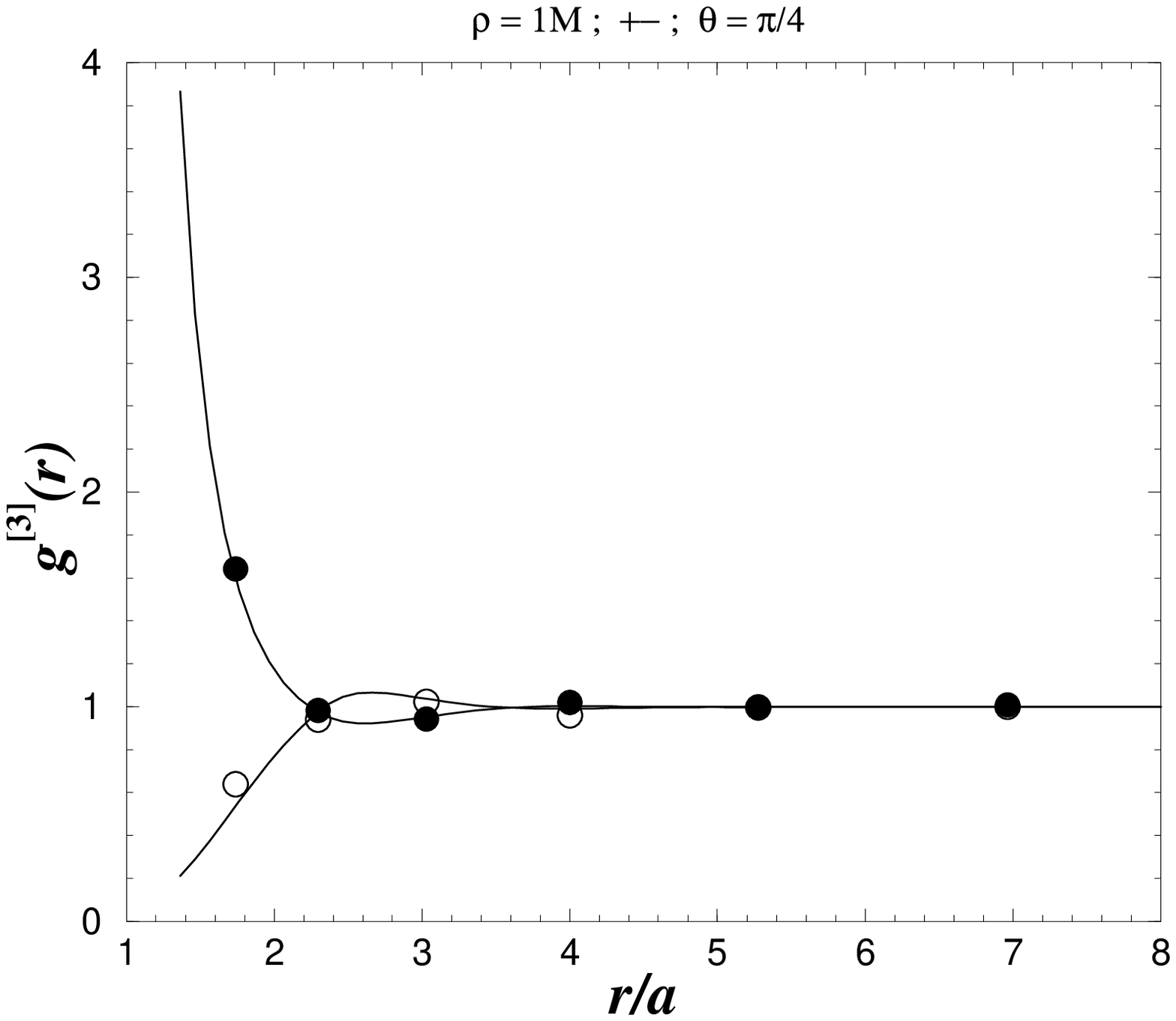}
\caption{Same as in Fig. \ref{fig.1M_90_+-} with $\theta =\pi /4$.
The MD results are shown in filled and open circles for
$g_{+--}^{[3]}(r,\pi /4;a)$ and $g_{+-+}^{[3]}(r,\pi /4;a)$
respectively.} \label{fig.1M_45_+-}
\end{figure}

Results for $\theta =\pi /4$ are shown in Fig. \ref{fig.1M_45_+-}.
We have again a very satisfactory agreement between theory and
simulation.

The 3D plot of $g_{+--}^{[3]}(r,\theta ;\tau
=a)=g_{+-+}^{[3]}(r,\pi -\theta ;\tau =a)$ obtained by TPE-HNC/MSA
is sketched in Fig. \ref{fig.plot3D_1M_+-}. The maximum and
minimum are located at $\theta =0$ and $\theta =\pi $ at dumbbell
contact, which implies a high probability of a line quadruplet
configuration. However, if we look at Fig.\ref{fig.1M_90_+-}, it
implies that, although with a lower probability, positive or
negative ions are adsorbed around the center of the antisymmetric
dumbbell, at $\theta =\pi /2$. Again, we observe oscillations at
any $\theta $ angle, and we checked that it is the case for any
$\tau $. Hence, charge reversal is also present, implying that
more than two ions are adsorbed to the dumbbell. Thus, probably
compact clusters more than line clusters are formed at this high
concentration. We will come back to this point later.

\subsubsection{Effective force}
\label{force_concentrated}

The effective mean force between two like charges [$F_{++}(r)$]
and that between two opposite charges [$F_{+-}(r)$] as a function
of their mutual separation $r$ are depicted in Fig.
\ref{fig.force_1M}, in reduced units of ${\displaystyle
\frac{k_{B}T}{\ell_{B}}}$. As expected, theories (TPE-HNC/MSA and
HNC/MSA) and simulation are in very good agreement for both forces
$F_{++}(r)$ and $F_{+-}(r)$.

An interesting feature is the kink in $F_{++}(r)$ occurring at
$r=2a$, that is somewhat less marked, however present, on the
simulation plot (due to the softness of the ions and also the
lower radial resolution there). This jump in the first derivative,
$F_{++}^{\prime }(r)$, at $r=2a$ is not an artifact of the theory
(or the simulation) but a true physical feature. This effect is
due to excluded volume correlations and, in much lesser degree, to
electrostatic correlations. It is clear that, at $r=2a$, the
configuration consisting of a counter-ion lying exactly between
two co-ions (i.e., $+-+$) is energetically very favorable (see
Fig.~\ref{fig.plot3D_1M_++}b). This implies the formation of ion
complexes, in qualitative agreement with Caillol and Weiss
\cite{caillolJCP_1995} and Yan and de Pablo
\cite{depabloJCP_2001}. When $r>2a$ (more precisely $r\rightarrow
2a^{+}$), the presence of an in-between counter-ion leads to a
relatively strong resistance, on the level of the depletion force,
upon approaching the two co-ions. On the other hand, when $r<2a$
(more precisely $r\rightarrow 2a^{-} $ ), the absence of an
in-between counterion leads to an easier approach (on the level of
the depletion force) of the two co-ions. These mechanisms, explain
(i) the discontinuity of $F_{++}^{\prime }(r)$ at $2a$ and (ii)
the fact that $|F_{++}^{\prime }(r\rightarrow
2a^{-})|<|F_{++}^{\prime
}(r\rightarrow 2a^{+})|$. This effect should also be observed in \emph{%
neutral} hard spheres systems at sufficiently high density, and in
the interaction between two macroions.

As far as the force $F_{+-}(r)$ is concerned, this kind of
discontinuity in the derivative is absent or nearly undetectable.
This is due to the fact that, at $r=2a$, the probability of
finding the configuration consisting of an ion between two
oppositely ions (i.e., $+--$) is considerably smaller
compared to that obtained with the configuration $+-+$. $F_{+-}(r)<0$ and $%
F_{++}(r)>0$ are of the same order of magnitude and indicate, of
course, that the ($+-$) configuration is of high probability,
whereas the ($++$) is of low probability.

By definition $\ln g_{\beta \gamma i}^{[3]}(r,\theta ;\tau )\equiv
$ $-w_{\beta \gamma i}(r,\theta ,\tau )/k_{B}T$. Hence, $\rho
_{i}g_{\beta \gamma i}^{[3]}(r,\theta ;\tau )$ gives the
probability of finding an ion of species $i$, at a certain
position $(r,\theta )$, from a dumbbell made of two ions of
species $\beta $ and $\gamma $, located at a distance $\tau $,
from each other. $w_{\beta \gamma i}(r,\theta ,\tau )$ is the
potential of mean force between the ion $i$ and the dumbbell. The
mean internal energy of an ideal gas per particle is $3k_{B}T/2$.
Hence, $-W_{0}\equiv -w_{++-}(r,\theta ,\tau )/[k_{B}T]>3/2$, for
a plus-plus pair, i.e. $g_{++-}^{[3]}(r,\theta ;\tau )>4.48$,
implies that a counter-ion next to a like-charged dumbbell has an
adsorption energy larger that its thermal energy, and thus it is
tightly attached to the dumbbell. For a 1M electrolyte,
$-W_{0}>3/2$ implies $\rho_{-}g_{++-}^{[3]}(r,\theta ;\tau )>4.48$
M. In Fig.\ref{fig.plot3D_1M_++}a, the peak is for
$\rho_{-}g_{++-}^{[3]}(r_{0},\theta =\pi /2;\tau =a)=50M \gg
4.48M$, i.e., $-w_{++-}(r,\theta =\pi /2,\tau =a)/k_{B}T=3.9>
1.5$. On the other hand, at $\theta =\pi $, $\rho
_{-}g_{++-}^{[3]}(r_{0},\theta =\pi ;\tau =a)=$5M, i.e.,
$-w_{++-}(r,\theta =\pi ,\tau =a)/k_{B}T$=1.6$\simeq $1.5.
Therefore, a negative ion will be strongly attached to the
positive ions pair (at $\theta =\pi /2$). A simple calculation
shows that the unscreened attractive electrostatic energy of a
second negative ion to the ($++-$) ion complex decreases to around
$50\%$ of the attractive energy of the positive ion pair to the
first negative ion. Hence, a second adsorbed ion, at $\theta =\pi
/2$, seems likely. Thus, Fig. \ref{fig.plot3D_1M_++}a suggest a
quadruplet structure, where the two counter-ions are at $\theta
=\pi /2.$ Clearly, more than two counter-ions are adsorbed, since
the dumbbell charge is overcompensated, i.e., there are
concentration profile oscillations. The adsorption of these
additional counter-ions is due to the short range correlations,
i.e., ions next to the dumbbell feel a net force toward it due to
the uneven collisions from bulk ions, and is an entropic effect,
beyond the ideal gas entropy. This effect is larger, the larger
the electrolyte concentration, and it will \textit{not} be present
in a point ion electrolyte. Because the attractive potential of
mean force is very high, this compact ion complex structure is
very stable, although very unlikely, because $F_{++}(r)>0$ (see
Fig.~\ref{fig.force_1M}). However, for $2a<\tau<3a$, a $(+-+)$
configuration is very likely. Hence, this indicates that there are
several mechanisms for the formation of ion complexes.

\begin{figure}
\includegraphics[width=12.0cm]{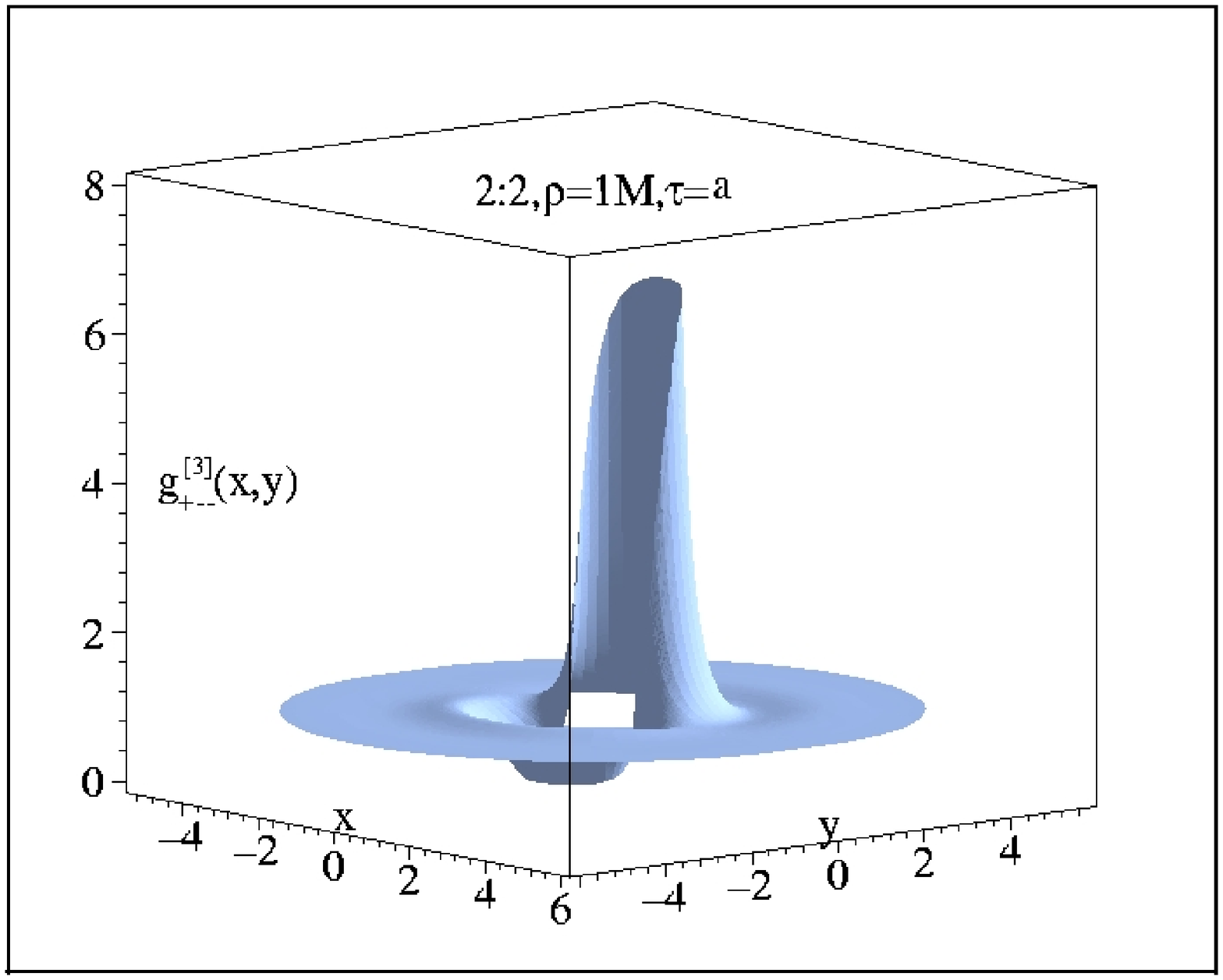}
\caption{3D representation (in Cartesian coordinates) of $%
g_{+--}^{[3]}(r,\theta; a)$ obtained by TPE-HNC/MSA for the same
fluid parameters as in Figs.\ref{fig.1M_90_+-} and
\ref{fig.1M_45_+-}. The dumbbell axis is parallel to $y$ axis.}
\label{fig.plot3D_1M_+-}
\end{figure}

\begin{figure}
\includegraphics[width=12.0cm]{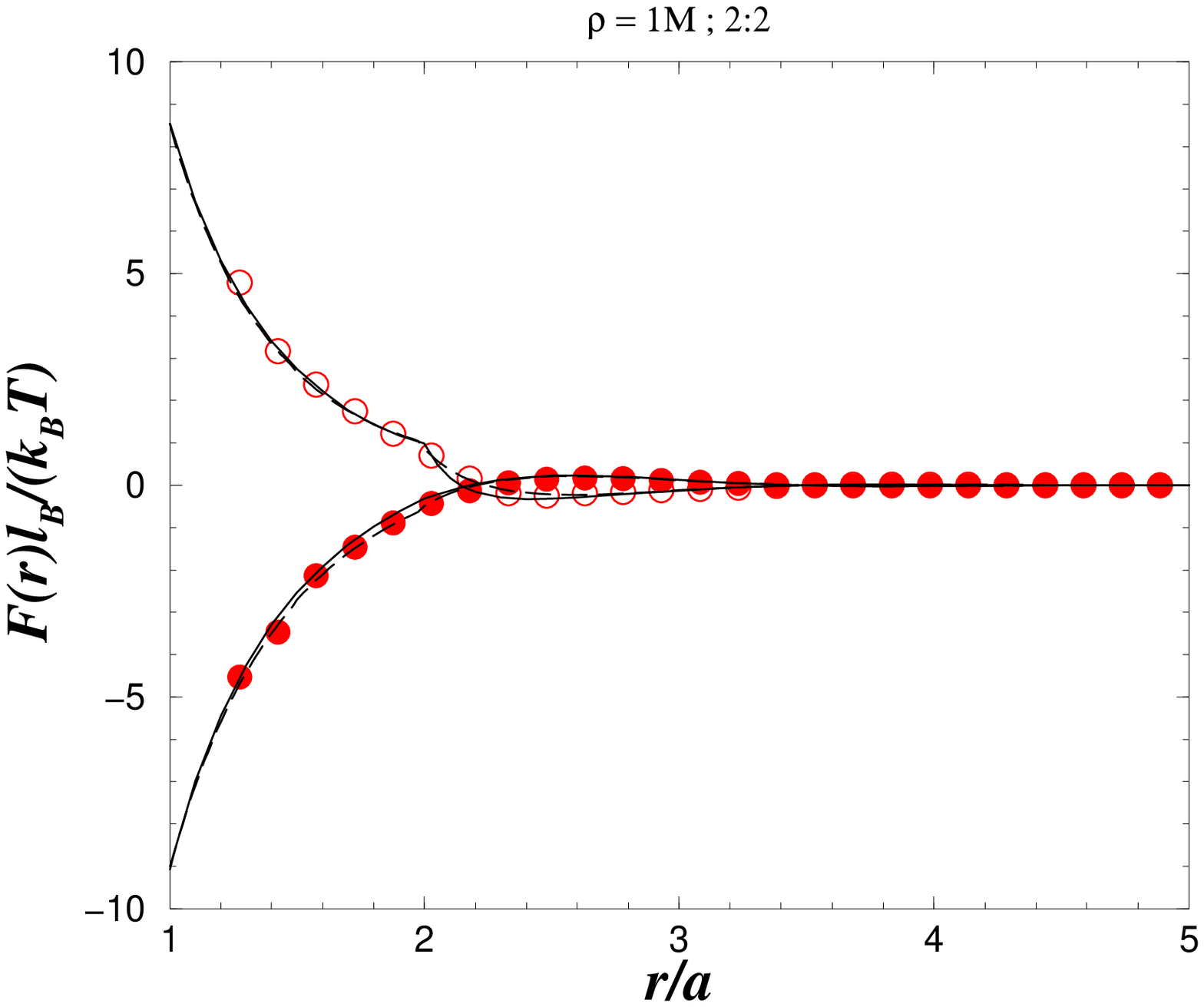}
\caption{Effective forces between two like charges [$F_{++}(r)>0$]
and two opposite charges [$F_{+-}(r)<0$] as a function of their
separation $r$, in reduced units of ${\displaystyle
\frac{k_{B}T}{\ell_{B}}}$, with $\rho =1$M and $z=2$. The solid
lines represent the results from TPE-HNC/MSA and the dashed lines
are the results from HNC/MSA. The MD results are shown in filled
and open circles for $F_{+-}(r)$ and $F_{++}(r)$, respectively.}
\label{fig.force_1M}
\end{figure}

For the plus-minus pair, in Fig \ref{fig.plot3D_1M_+-} the peak is
for $\rho _{+}g_{+--}^{[3]}(r_{0},\theta =0;\tau =a)=6.8M>4.48M$,
i.e., $ -w_{+--}(r,\theta =0,\tau =a)/k_{B}T=1.9>1.5$. and from
Fig. \ref{fig.1M_90_+-}, $-w_{+--}(r,\theta =\pi /2,\tau
=a)/k_{B}T=0.18 \ll 1.5$. Hence, for an unlike charged dumbbell,
we expect an aligned stable quadruplet configuration (because of
symmetry), due to energy arguments. However$,$ due to entropic
effects more counter-ions are adsorbed into the dumbbell,
producing charge reversal, as can be seen from the oscillations of
Fig.\ref {fig.plot3D_1M_+-}. These additional ions are delocalized
around the dumbbell, hence, generating compact ion complexes
because $F_{+-}(r)<0$ (see Fig.~\ref{fig.force_1M}), this
configuration is very likely.

\subsection{Dilute case}

In this section, we study a dilute, divalent electrolyte
($\rho=0.005$M and $z=2$). To the best of our knowledge {\em all
of the known liquid theories fail to describe the RPM behavior
under these conditions} \cite{lozada83,sjur83}. Hence, the study
of low concentrated solutions of multivalent ions represents a
strong test case for a liquid theory. In addition, for the RPM
electrolyte we are on the low concentration side of the phase
diagram.

\subsubsection{Three particle correlation function}

\paragraph{Symmetric dumbbell}

\begin{figure}
\includegraphics[width=12.0cm]{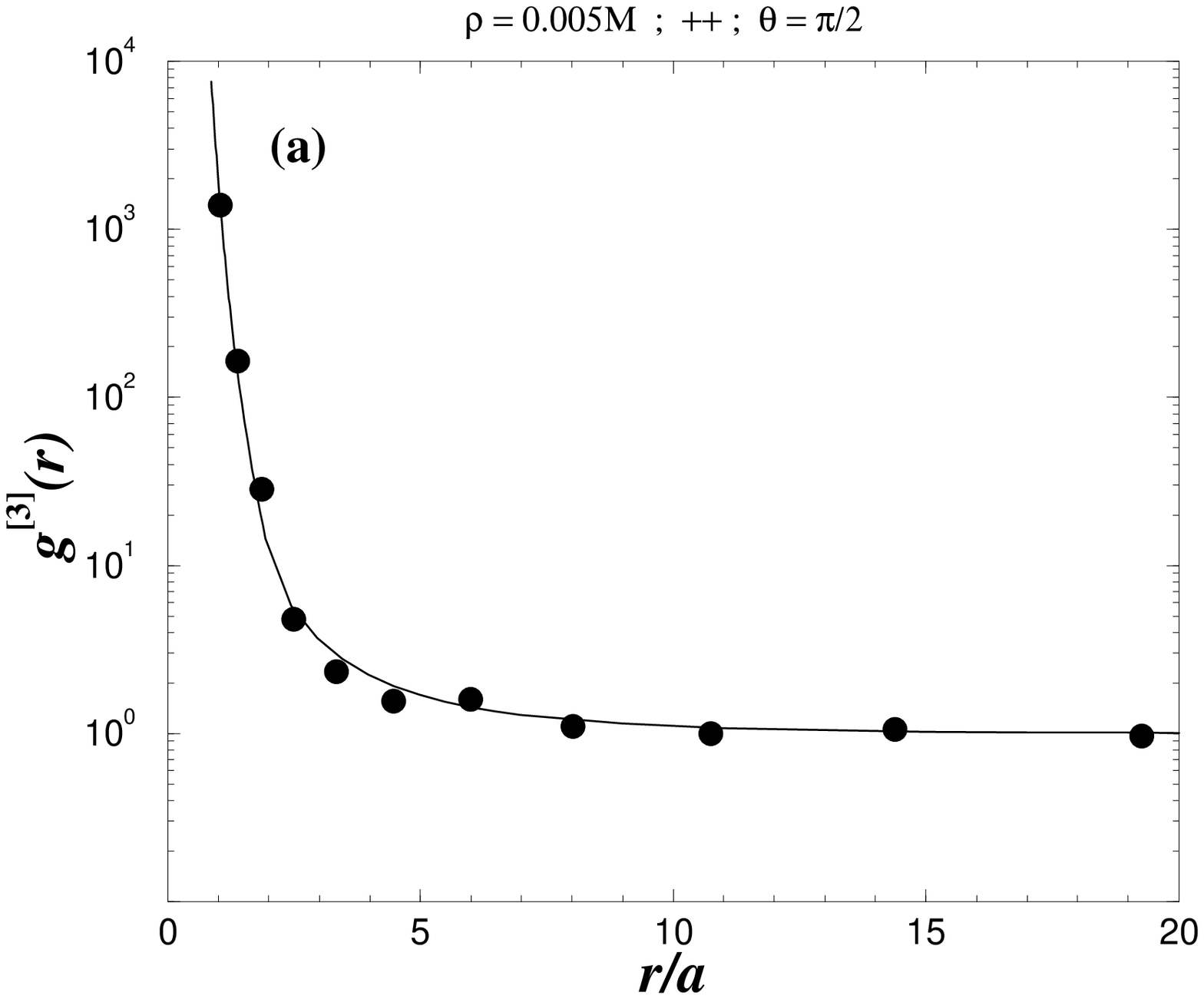} %
\includegraphics[width=12.0cm]{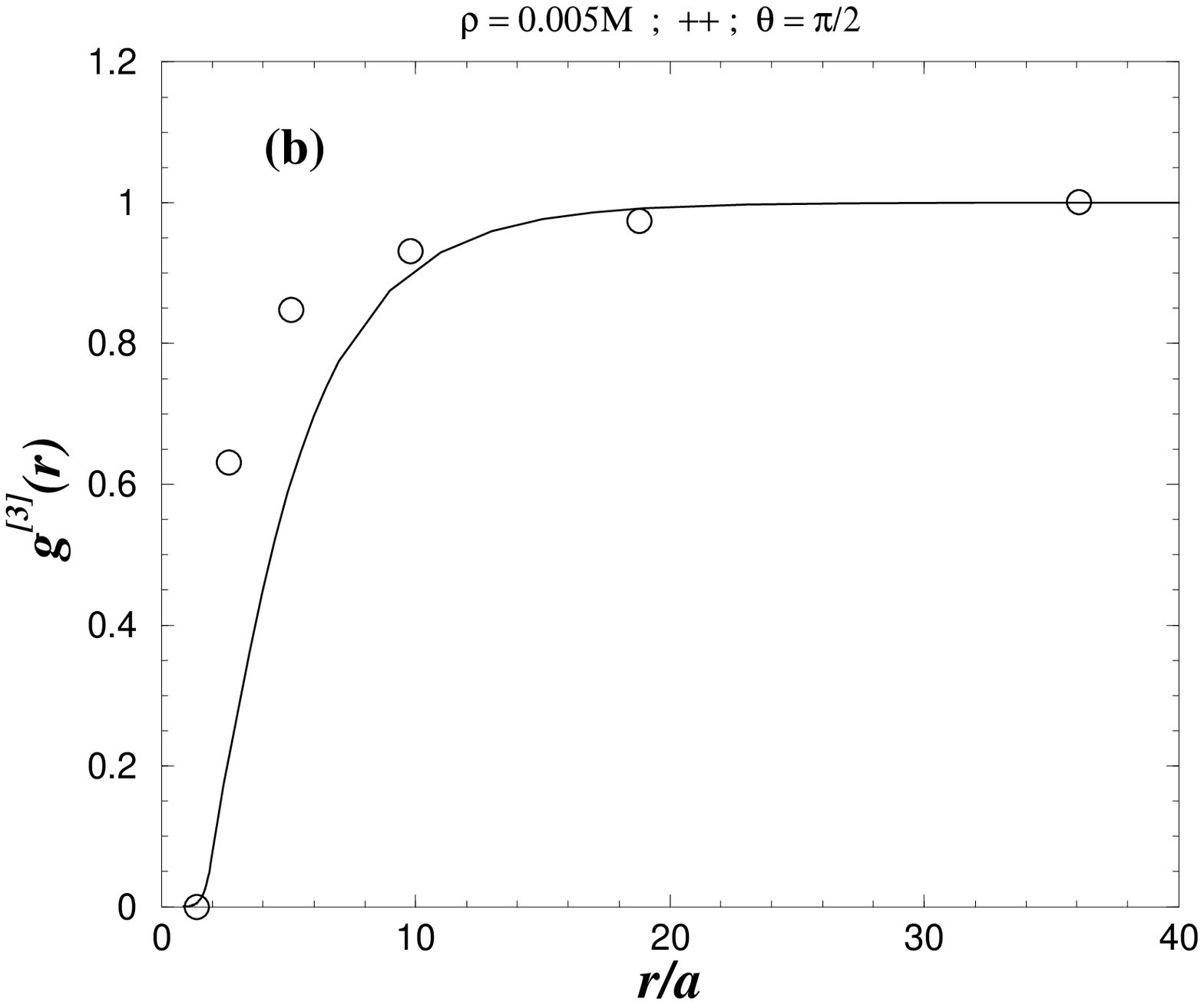}
\caption{Three particle distribution function with $\rho=0.005$M
and $z=2$: (a) $g_{++-}^{[3]}(r,\pi/2; a)$ and (b)
$g_{+++}^{[3]}(r,\pi/2;a)$. The solid lines represent the results
from TPE-HNC/MSA. The MD results are shown
in filled and open circles for (a) $g_{+--}^{[3]}(r,\pi/2; a)$ and (b) $%
g_{+-+}^{[3]}(r,\pi/2;a)$, respectively.} \label{fig.005M_90_++}
\end{figure}

Figures \ref{fig.005M_90_++}(a) and \ref{fig.005M_90_++}(b) show a
comparison between TPE-HNC/MSA and MD results for $g_{++-}^{[3]}(r,\pi /2;a)$ and $%
g_{+++}^{[3]}(r,\pi /2;a)$, respectively. One can see that the
electrical double layer is wider than in the concentrated case
i.e., the correlations are long ranged. For $g_{++-}^{[3]}(r,\pi
/2;a)$ [see Fig. \ref {fig.005M_90_++}(a)], TPE-HNC/MSA and MD
results show quantitative agreement, even near contact. For the
$g_{+++}^{[3]}(r,\theta =\pi /2;a)$ [see Fig. \ref
{fig.005M_90_++}(b)], a qualitative agreement between TPE-HNC/MSA
and MD results is found.

\begin{figure}
\includegraphics[width=12.0cm]{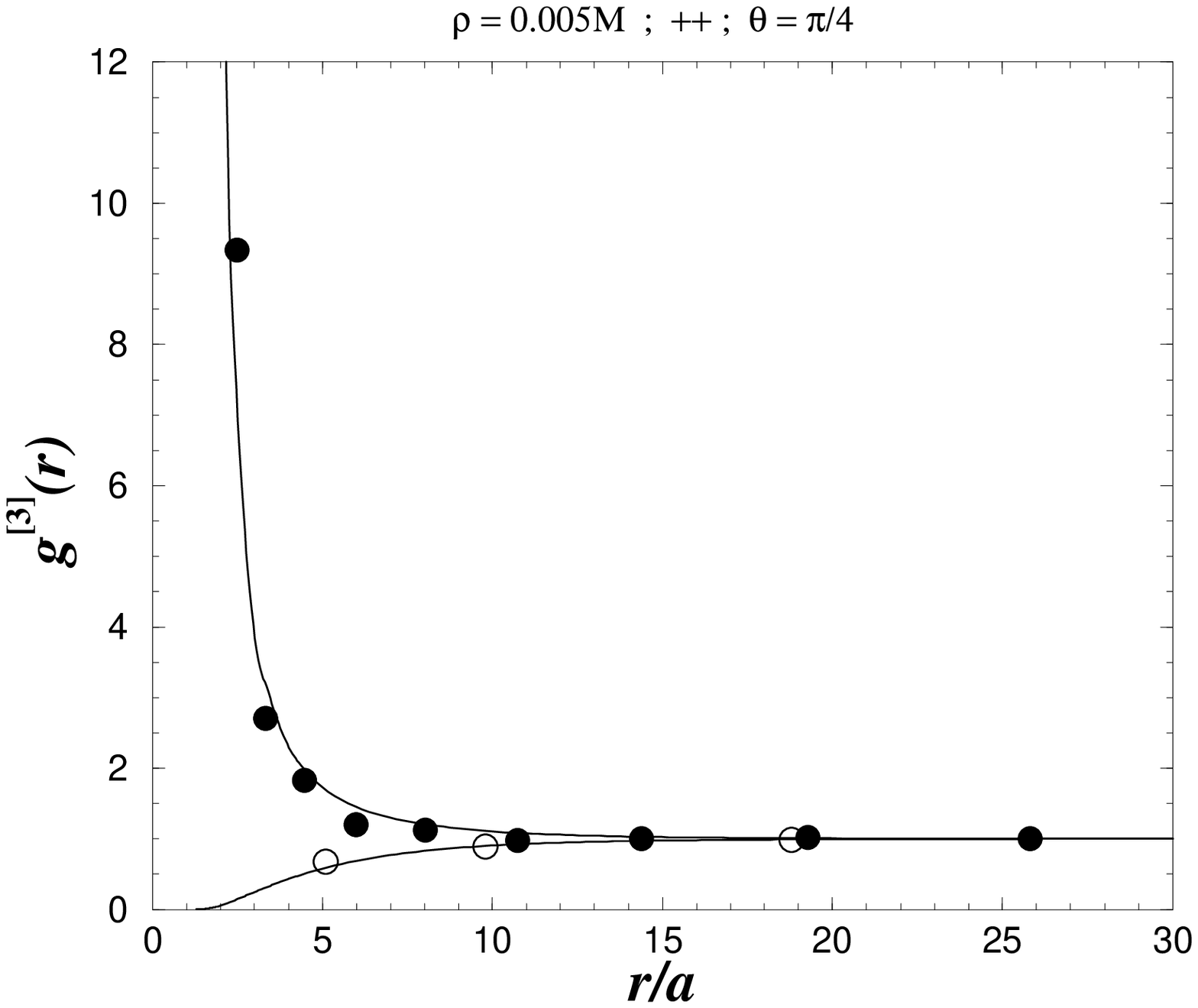}
\caption{Same as in Fig. \ref{fig.005M_90_++} with $\theta=\pi/4$.
The MD results are shown in filled and open circles for
$g_{++-}^{[3]}(r,\pi/2; a)$ and $g_{+++}^{[3]}(r,\pi/2;a)$,
respectively.} \label{fig.005M_45_++}
\end{figure}

At $\theta =\pi /4$ (see Fig.~\ref{fig.005M_45_++}) it is found
that the contact value of $g^{[\mathrm{{3}]}}_{++-}(r,\pi/4;a )$
(about 500) is much smaller, of two orders of magnitude, than that
at $\theta=\pi/2$. This can
be explained in terms of the electric field produced by the dumbbell at $%
\theta=\pi/2$ which is considerably stronger than at
$\theta=\pi/4$.

\begin{figure}
\includegraphics[width=12.0cm]{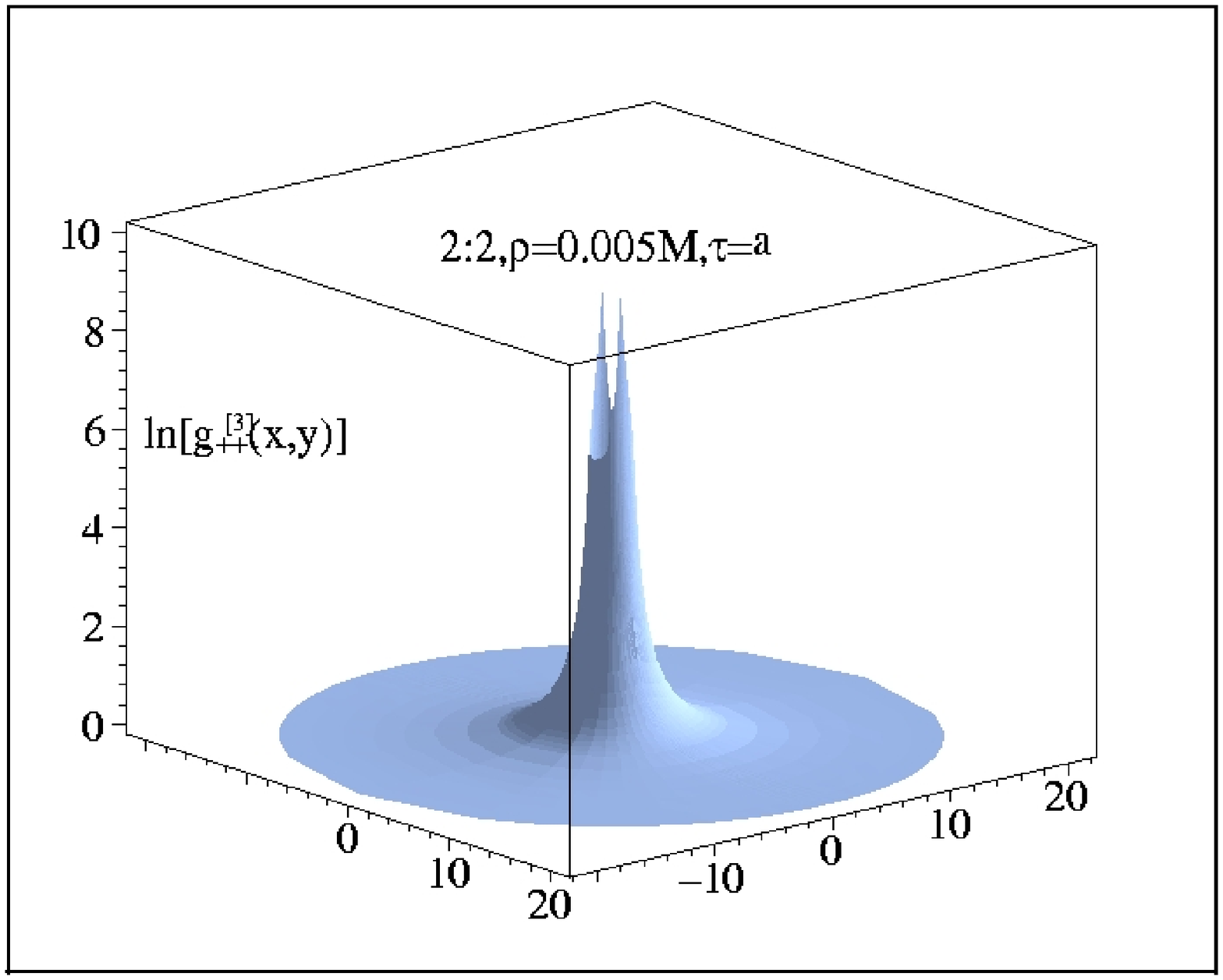}
\includegraphics[width=12.0cm]{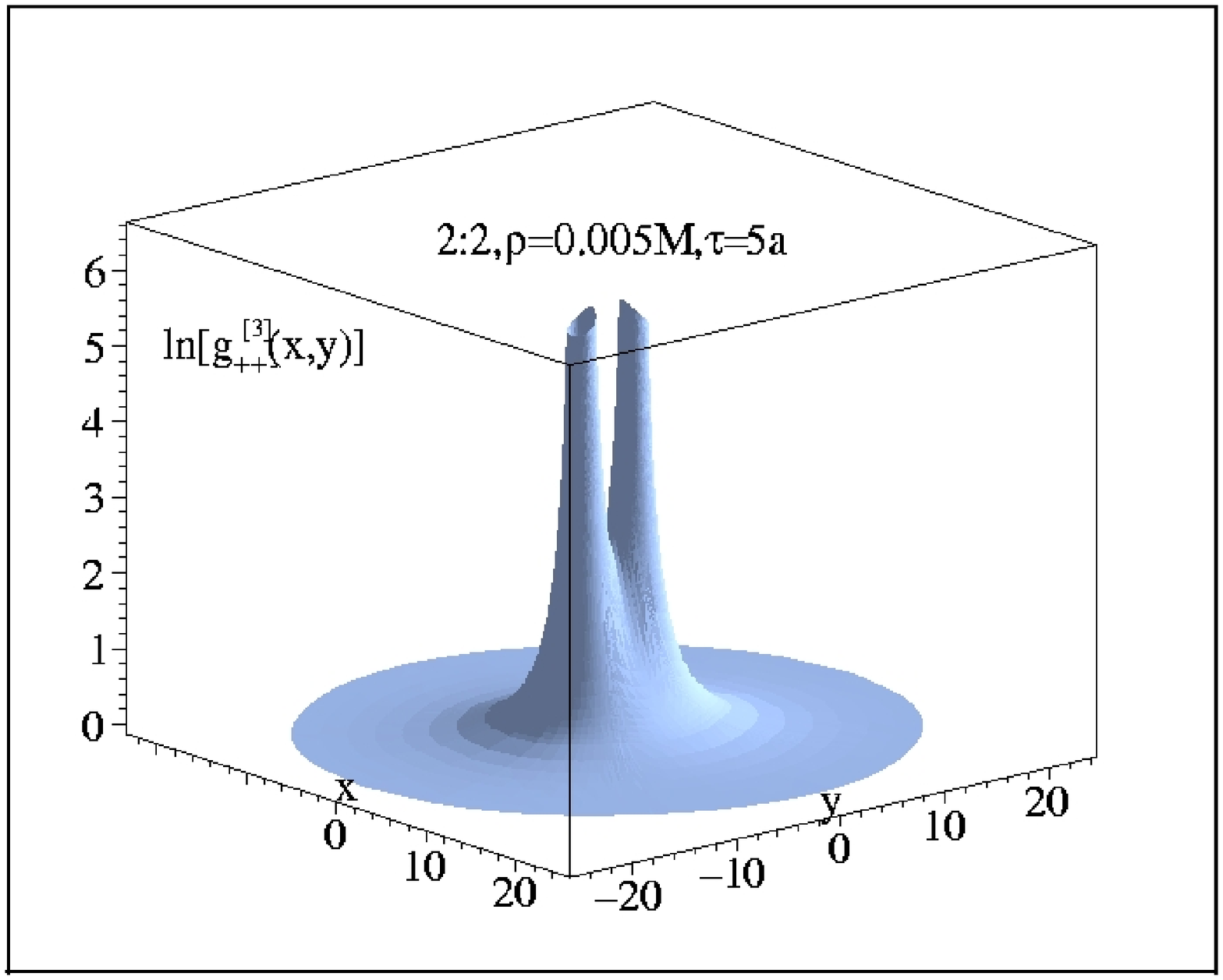}
\caption{3D representation (in Cartesian coordinates) of $%
g_{++-}^{[3]}(r,\theta ;\tau )$ obtained by TPE-HNC/MSA for the
same fluid parameters as in Figs.~\ref{fig.005M_90_++} and
\ref{fig.005M_45_++}. The dumbbell axis is parallel to $y$ axis.
(a) $\tau =a$ (b) $\tau =5a$. } \label{fig.plot3D_005M_++}
\end{figure}

The 3D plot of $\ln (g_{++-}^{[3]}(r,\theta ;\tau ))$ can be found
in Fig.~ \ref{fig.plot3D_005M_++}. For $\tau =a$ [see
Fig.\ref{fig.plot3D_005M_++} (a)], it is observed a strong
variation of the distribution function close to the dumbbell (at
the surface of closest approach). As expected, the maximum of
$g_{++-}^{[3]}(r,\theta ;a)$ is at $\theta =\pi /2$. On the other
hand, at $\tau =5a$ (see Fig.~\ref{fig.plot3D_005M_++}(b)), the
angular variation of $g_{++-}^{[3]}(r,\theta ;5a)$ (near contact)
around one ion of the dumbbell is not as peaked as in
$g_{++-}^{[3]}(r,\theta ;a)$. However, the dumbbell ions are still
correlated, i.e., their electrical double layers are strongly
\emph{overlapped} although $\tau =5a$. We had to go up to $\tau
=60a$ (not shown) to cancel the overlapping of the electrical
double layers of the dumbbell ions. At low salt concentration
there is a longer range penetration of the ions electrical field
into the fluid, and hence charge correlations are of longer range.
For low salt concentration the role of higher order diagrams is
more important. It is observed that {\em for this low
concentration case, there are no oscillations} in the counterion
concentration profiles. Hence, no charge reversal is present and,
thus, the formation of a simple more complex ionic configurations,
beyond a quadruplet formation, is not supported by our results.
Because of the very large value of $W_{0}\approx 8.6$, the
adsorption of two counter-ions to the like-charged dumbbell (at
$\theta =\pi /2$) is much larger than for the equivalent situation
for the concentrated case, where $W_{0}\approx 3.9.$ Hence, for
the dilute case the quadruplet is more stable, but even less
probable due to the lower concentration.

\paragraph{Antisymmetric dumbbell}

\begin{figure}
\includegraphics[width=12.0cm]{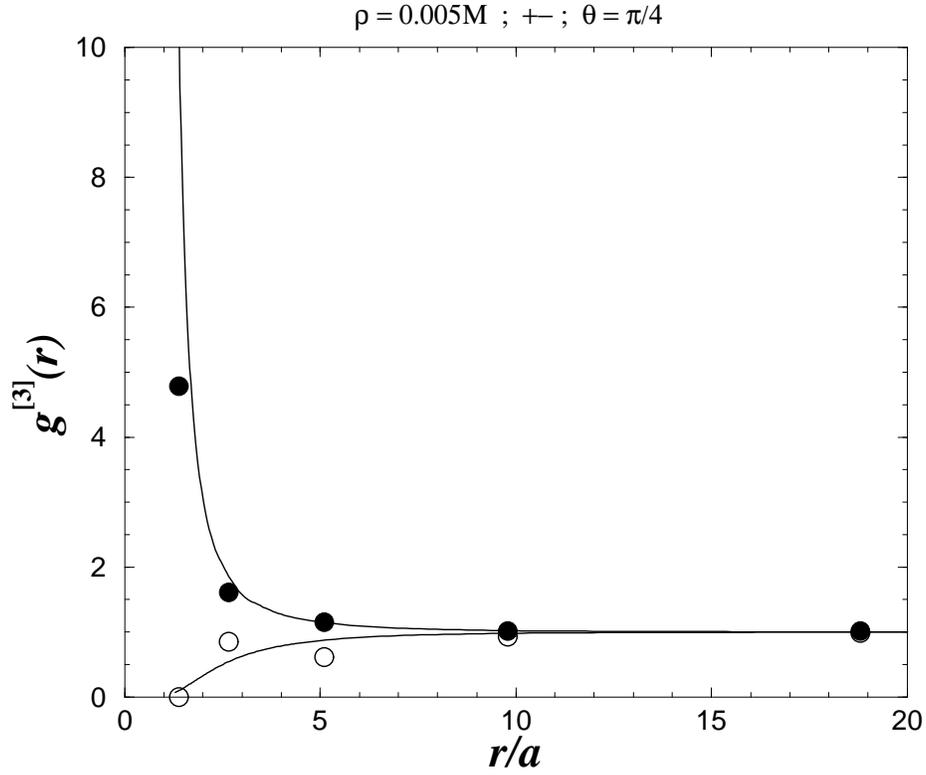}
\caption{Three particle distribution function
$g_{+-i}^{[3]}(r,\pi/4;a)$ with $\rho=0.005$M and $z=2$. The solid
lines represent the results from TPE-HNC/MSA. The MD results are
shown in filled and open circles for $ g_{+--}^{[3]}(r,\pi/4;a)$
and $g_{+-+}^{[3]}(r,\pi/4;a )$, respectively.}
\label{fig.005M_45_+-}
\end{figure}

We now consider the three particle distribution function where the
dumbbell is made up of two opposite ions, and for the same fluid
parameters as in
Figs. \ref{fig.005M_90_++} and \ref{fig.005M_45_++}. Only the case of $%
\theta =\pi /4$ is shown (see Fig.~\ref{fig.005M_45_+-}), given that for $%
\theta =\pi /2$ the electrical field is zero and since the
electrolyte
concentration is very low we have $g_{+--}^{[3]}(r,\theta =\pi %
/2)=g_{+-+}^{[3]}(r,\theta =\pi /2)\approx 1$. This is in contrast
with the 1M electrolyte result of Fig. \ref{fig.1M_90_+-}. For
$\theta =\pi /2$ the same good agreement is found between
TPE-HNC/MSA and MD as that in Fig. \ref {fig.005M_45_++}.

\begin{figure}
\includegraphics[width=12.0cm]{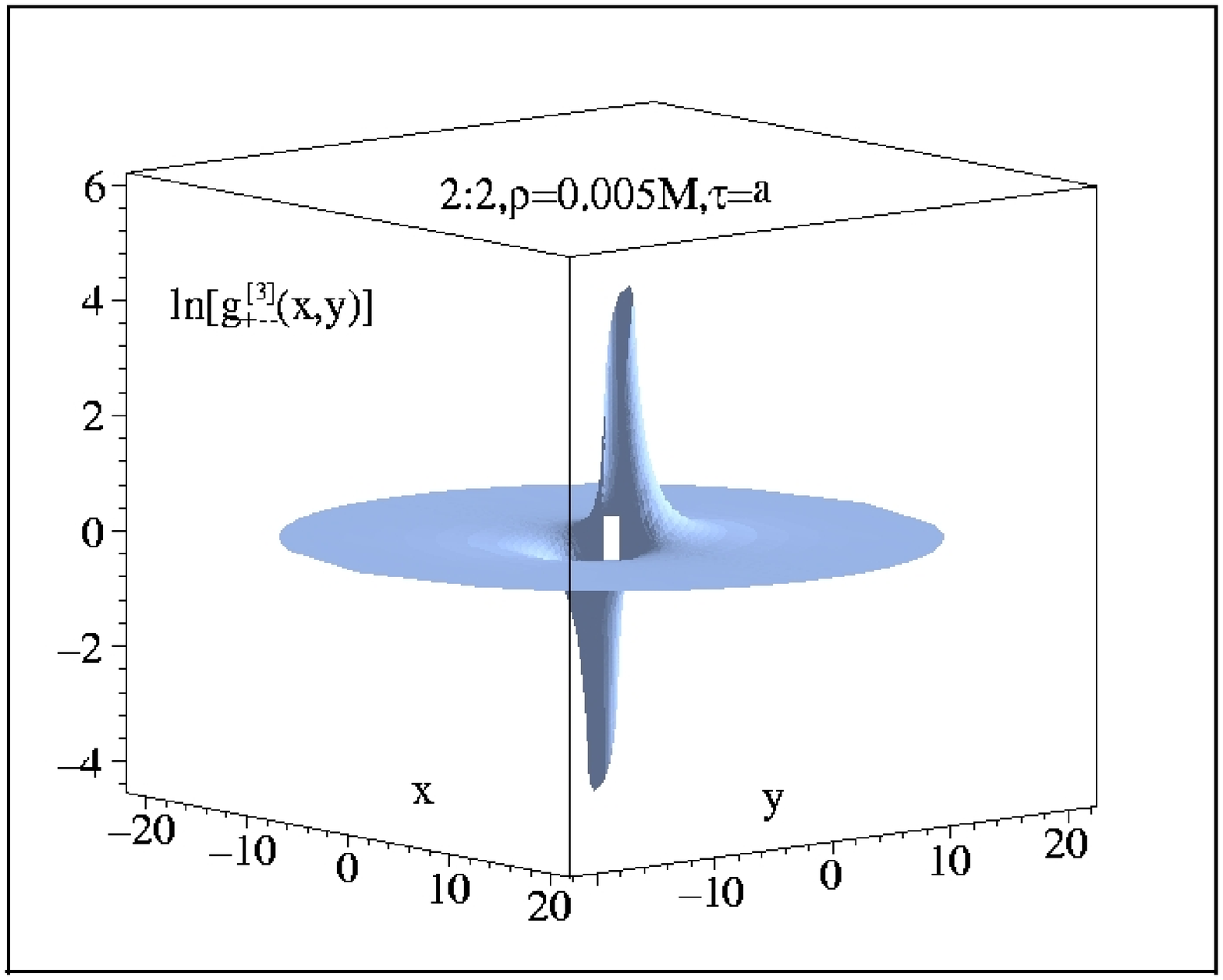}
\caption{3D representation (in Cartesian coordinates) of $%
g_{+--}^{[3]}(r,\theta ;a)$ obtained by TPE-HNC/MSA for the same
fluid parameters as in Fig.~\ref{fig.005M_45_+-}. The dumbbell
axis is parallel to $y$ axis.} \label{fig.plot3D_005M_+-}
\end{figure}

The 3D plot of $\ln g_{+--}^{[3]}(r,\theta ;\tau =a)=\ln
g_{+-+}^{[3]}(r,\pi -\theta ;\tau =a)$ which is the potential of
mean force is sketched in Fig.~ \ref{fig.plot3D_005M_+-}. This
function is quasi center-symmetric with respect to the dumbbell
center. This feature is due to (i) the symmetry of the
electrostatic correlations and (ii) the fact that the contribution
of the excluded volume correlations (at such low density) is
negligible compared to that in the concentrated case. As expected
the function is strongly peaked at $\theta =0$. For the
concentrated case (not shown) the asymmetry is higher. The
important result shown in this figure is the formation of a
stronger ($+-+-)$ line quadruplet, than for the concentrated case,
since here $-w_{+--}(r=3/2a,\theta =0,\tau
=a)/k_{B}T=-w_{+-+}(r=3/2a,\theta =\pi,\tau =a)/k_{B}T\approx 4.2>
1.5$, which is much higher than that for the corresponding
concentrated case ($-w_{+--}(r=3/2a,\theta=0,\tau=a)/k_BT\approx
1.6$). Hence, this line quadruplet structure would be more stable.
This result suggests that quadruplets, if present, would be in a
linear configuration more than in compact quadruplets structures,
in disagreement with Fig.~4 of Yan and de Pablo
\cite{depabloJCP_2001,depabloCMT}.

\subsubsection{Effective force}

\label{force_dilute}

\begin{figure}
\includegraphics[width=12.0cm]{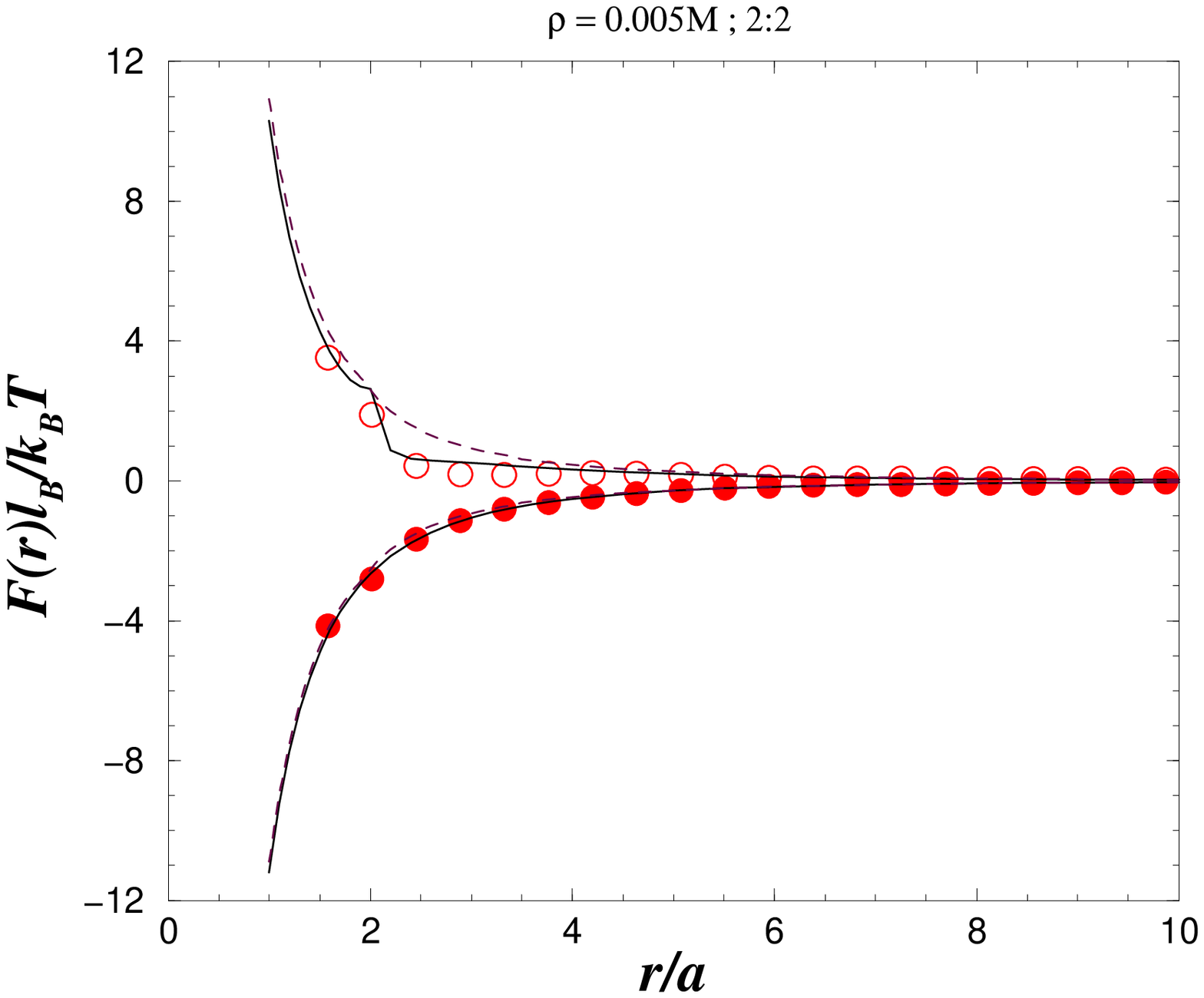}
\caption{Effective forces between two like charges [$F_{++}(r)>0$]
and two opposite charges [$F_{+-}(r)<0$] as a function of their
separation $r$, in reduced units of ${\displaystyle
\frac{k_{B}T}{\ell_{B}}}$, with $\rho =0.005$M and $z=2$. The
solid lines represent the results from TPE-HNC/MSA and the dashed
lines are the results from HNC/MSA. The MD results are shown in
filled and open circles for $F_{+-}(r)$ and $F_{++}(r)$ ,
respectively.} \label{fig.force_005M}
\end{figure}

The effective mean force between two like charges [$F_{++}(r)$]
and that between two opposite charges [$F_{+-}(r)$] as a function
of their mutual separation $r$ can be found in Fig.
\ref{fig.force_005M}, in reduced units of ${\displaystyle
\frac{k_{B}T}{\ell_{B}}}$. Concerning $F_{+-}(r)$, theories
(TPE-HNC/MSA and HNC/MSA) and simulation are in quantitative
agreement.

As pointed out above, for $F_{++}(r)$ in the concentrated case,
the derivative $F_{++}^{\prime }(r)$ is again discontinuous at
$r=2a$. The same mechanism proposed for the concentrated case (see
Sec.~\ref {force_concentrated}) applies here. This important
feature is not captured by HNC/MSA, proving the qualitative
improvement by using the TPE method. This better description
steams from proper inclusion of long ranged correlations. Finally,
we have a good quantitative agreement between TPE-HNC/MSA and MD.
In comparison of Fig. \ref{fig.force_005M}, with that for the
concentrated case, Fig. \ref{fig.force_1M}, two important
differences are observed: The intensity and the range of the force
is larger for the dilute case, implying that the electrical field
is less screened. Also for low concentration $F_{++}>0$, i. e., it
is always repulsive, whereas in the concentrated regime, for some
interval of $\tau$, $F_{++}$ is negative, implying an attraction
and hence different ion-complexes mechanisms. In addition, one can
expect that for a certain combination of temperature, solvent
dielectric constant, and salt valence and low concentration, one
can find a phase transition, in which associated ions and free
ions coexist: {\em single ions, ion pairs and quadruplets}. Hence,
from Figs.~\ref{fig.plot3D_1M_+-} and \ref{fig.force_1M}, for the
concentrated case, and Figs.~\ref{fig.plot3D_005M_+-} and
\ref{fig.force_005M}, for the dilute case, we conclude that linear
ion complexes are likely to be formed. At low concentration,
dumbbells ($+-$) and line quadruplets ($+-+-$) are very likely to
be formed, whereas at high concentration larger complexes than
quadruplets are formed. This is in qualitative agreement with the
predictions of Caillol and Weiss \cite{caillolJCP_1995}.

\section{Conclusions}

\label{sec.Conclusions}

We have investigated the structure of 1:1 and 2:2 RPM electrolytes
by means of integral equations and MD simulations. Using the three
point extension to
the HNC/MSA theory, the conditional three particle distribution function, $%
g^{[3]}(r,\theta;\tau)$, was computed and compared with that
obtained by MD. {Although it is not shown, for the 1:1 electrolyte
the quantitative agreement between TPE-HNC/MSA and MD is
excellent}. For the 2:2 electrolyte, we explicitly report here
results for two typical concentrations: (i) the concentrated case
($\rho=1$M) and (ii) the dilute case ($\rho=0.005$M).

As far as the concentrated case concerns, it was found that $%
g^{[3]}(r,\theta)$ always presents oscillations. The detailed
comparison between TPE-HNC/MSA and MD, carried at fixed separation
$\tau=a$ (between the two constitutive ions of the dumbbell),
shows an excellent qualitative and/or quantitative agreement. This
is true for all values of $\tau$ (not shown).

On the level of the effective mean force between two ions, both,
TPE-HNC/MSA and HNC/MSA are in very good agreement with MD. This
is consistent with previous comparisons between HNC/MSA and Monte
Carlo results \cite{lozada83,degreve93}. Hence we can conclude
that the TPE-HNC/MSA method is also {suitable} to describe
concentrated electrolyte solutions. It is important to point out a
particular behavior in the effective force between like-charged
ions [$ F_{++}(r)$] observed at $r =2 a$, where an abrupt change
in its slope appears due to excluded volume correlations. This
behavior can not be directly seen in the pair distribution
function (for this value of $\rho$).

In the dilute regime, the analysis of the three particle
distribution function and the effective force shows the long range
nature of the correlations. For the three particle distribution
functions, we had to go up to a distance separation of $\tau
\approx 60a$, in order to uncorrelated the two constitutive
dumbbell ions. Again a good agreement for $g^{[3]}(r,\theta )$ is
found between TPE-HNC/MSA and MD, proving the robustness of the
TPE formalism. The study of the effective force reveals a
quantitative agreement for the force between two oppositely
charged ions, $F_{+-}(r)$, between TPE-HNC/MSA and MD, although
HNC/MSA is also very good. For the force $F_{++}(r)$ we again
remark the occurrence of an abrupt change in its slope at $\tau
=2a$, which is not predicted by HNC/MSA. On the other hand,
TPE-HNC/MSA and MD are in quantitative agreement, showing the
ability of TPE to take fairly well into account long range
correlations. It is precisely in this region of the ionic fluid
phase diagram, i.e., low concentration and high Coulombic
coupling, where all the other theories fail.

The TPE approach is a general formalism that improves existing
liquid theories, by including higher order diagrams in a
systematic, consistent way \cite{henderson92a}. Here we have shown
it to be successful for ionic fluids, in all the regions of the
RPM phase diagram, in particular in the region of low salt
concentration and high coulombic coupling.

In the high concentration regime, ion pairs tend to form aligned
quadruplets, i.e., ($-+-+$) structures are energetically favored.
However, because of short range correlations, other delocalized
ions are adsorbed to produce charge reversal of the unsymmetrical
ion dumbbell and hence the formation of larger complexes than
quadruplets is favored. In the low concentration regime, the
($-+-+$) aligned quadruplet structure is even more stable than for
the high concentration case. Hence, dumbbells and aligned
quadruplets are likely to be formed. No adsorption of additional
ions is present, since there are no oscillations in the
concentration profile and, hence, there is no charge reversal of
the dumbbell or higher multiploles. In the high concentration
regime charge reversal is present, whereas at low concentration
there is no charge reversal. Our results clearly indicate the
formation of ion pairs and complexes, in agreement with previous
theoretical predictions
\cite{fisher94,blumJSP_1995,blumJCP_1996,blumMP_2001,blumJCP_2002,stellJCP_1995,stellJPC_1996,stellJP_2000}
and simulation results \cite{panaPRE_2002a,depabloJCP_2001}. In
our theory we {\em do not} impose ion pairs, and could be useful
to explore RPM phase transitions, critical behavior and could
provide a means to understand the molecular mechanisms behind
fluids phase transitions.

\acknowledgments

FJA and MLC thank CONACYT (L007E and C086A) and NEGROMEX for their financial
support. RM gratefully acknowledges the support of \textit{Labotatoires
Europ\'eens Associ\'es} (LEA).

\appendix

\section{ Numerical Method}

\label{sec.NumMet}

\subsection{Finite Element Method}

\label{sec.FEM} 
\begin{figure}
\includegraphics[width=12.0cm]{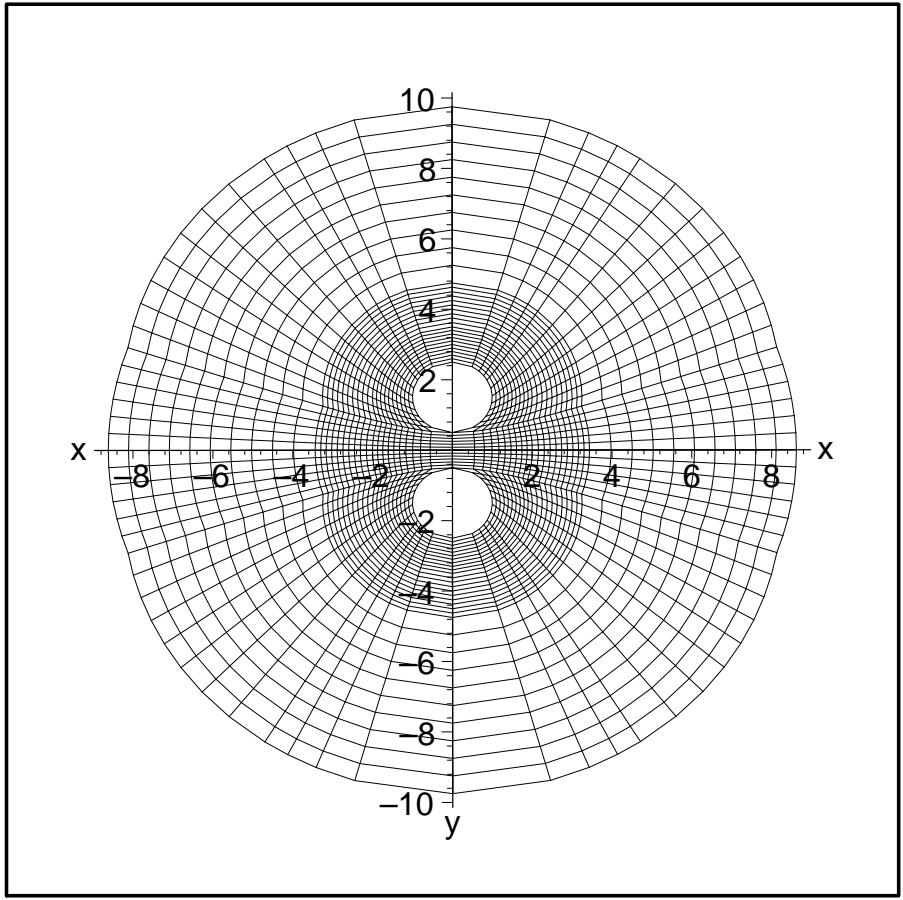}
\caption{An example of the grid (in Cartesian coordinates) used to
solve Eq.(\ref{numerical1}).} \label{cartesian}
\end{figure}

\begin{figure}
\includegraphics[width=12.0cm]{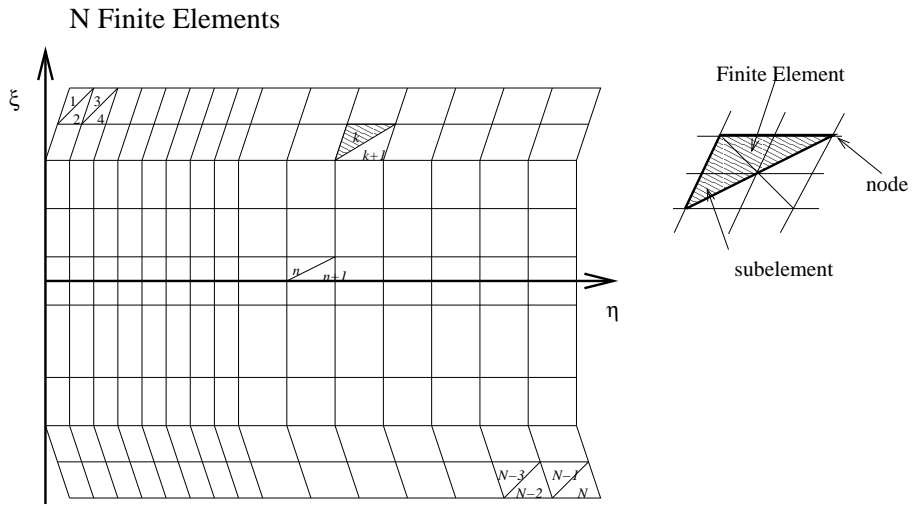}
\caption{The same grid as in in Fig. \ref{cartesian} but mapped into the $%
\eta-\xi$ plane. A triangular element and its 6 nodes are
represented.} \label{prolates}
\end{figure}

To solve the TPE-HNC/MSA equation, Eq. (\ref{tpe-hnc-msa1}),
it is necessary to use a numerical method, since an analytical solution is
not available. The finite element method (FEM) has been used in the past to
solve HNC/MSA equation in several geometries \cite{lozada90b,sanchez92,mier}
and it has been proved to be efficient. The general form of TPE-HNC/MSA
integral equation can be written as

\begin{eqnarray}  \label{numerical1}
g_{\alpha i}\left( \eta ,\xi \right) = \exp \left\{ \mathrm{M}
_{i}(\eta,\xi) +\int_{-1}^{1}\int_{\eta _{0}(\xi )}^{\infty
}\sum_{m=1}^{2} \rho _{m}  h_{\alpha m} \left( \eta ^{\prime },
\xi ^{\prime}\right) \mathrm{{F}\left( \eta ,\xi ,\eta ^{\prime
},\xi ^{\prime }\right) d\eta
^{\prime }d\xi ^{\prime }}\right\} , \\
\end{eqnarray}
where $h_{\alpha m}\left( \eta ,\xi \right)$ and $M_i(\eta,\xi)$ are
functions defined on a bidimensional domain $(\eta,\xi) \in
\left[-1,1\right] \times \left[ 1,\infty \right)$. Since $h_{\alpha m}\left(
\eta ,\xi \right) \neq 0$ only in a region close to the dumbbell, we solve
Eq.~(\ref{numerical1}) just in a finite domain. In the FEM \cite{zienkiewitz}
, the domain is divided into $N$ elements. Every element in a domain $A_{K}$
is divided into $L_{0}$ sub-elements. In prolate coordinates, the dumbbell
geometry of Fig. \ref{cartesian} is mapped into the geometry shown in Fig.
\ref{prolates}, where one of the $N$ triangular elements is shown.

In order to solve Eq. (\ref{numerical1}), the function $h_{\alpha m}\left(
\eta ,\xi \right)$ is expanded as a linear combination of a $L_{0}$ base
elements $\left\{ \phi_{i}^{K}\left( \eta ,\xi \right)
,i=1,...,L_{0}\right\} $, where $1\le K\le N$. These base functions are
defined in such a way that $\phi _{i}^{K}\left( \eta ,\xi \right)=0$ {if} $(
\eta ,\xi ) \notin A_{K}$. Furthermore the base functions are chosen so that
for a set of $L_{0}$ points $\left(\eta_{j},\xi _{j}\right)$ (which are
called nodes, see Fig. \ref{prolates}), they satisfy

\begin{equation}
\phi _{i}^{K}\left( \eta ,\xi \right)= \delta _{ij}, \hspace{0.2cm} \mathrm{%
\ with} \hspace{0.2cm} i,j=1,...,L_{0}.,
\end{equation}
with $\delta _{ij}$ being the Kronecker delta function. Hence,

\begin{equation}
h_{\alpha m}\left( \eta ,\xi \right) =\sum_{K=1}^{N}\sum_{l=1}^{L_{0}}\omega
_{ml}^{K}\phi _{l}^{K}\left( \eta ,\xi \right) \mathrm{{,}m=1,2}
\label{base0}
\end{equation}
where $\left\{ \omega _{ml}^{K},l=1,...,L_{0}\right\} $ are the $L_{0}$
coefficients of the $h_{\alpha m}$ in the $K$-th finite element. Thereby,
the coefficient $\omega _{mi}^{K}$ is the value of the function at the $i$
-th node, i.e.,

\begin{equation}
\omega _{mi}^{K}=h_{\alpha m}\left( \eta _{i},\xi _{i}\right) .
\end{equation}
It is useful to renumber $\phi _{l}\left( \eta ,\xi \right)$ and $\omega
_{ml}$, so that Eq. (\ref{base0}) can be rewritten as follows

\begin{equation}
h_{\alpha m}\left( \eta ,\xi \right) =\sum_{l=1}^{M_{0}}\omega _{ml}\phi
_{l}\left( \eta ,\xi \right) ,\mathrm{{\ }m=1,2}  \label{base1}
\end{equation}
where
\begin{equation}
\omega _{mi}=h_{\alpha m}\left( \eta _{i},\xi _{i}\right)= g_{\alpha
m}\left( \eta _{i},\xi _{i}\right)-1,  \label{coefficient1}
\end{equation}
with $M_{0}=N\times L_{0}$, and the superscript $K$ has been omitted. By
substituting Eq. (\ref{base1}) into Eq. (\ref{numerical1}), we get

\begin{equation}
g_{\alpha i}( \eta ,\xi ) =\exp \left\{ \mathrm{{M}_{i}( \eta ,\xi )
+\sum_{l=1}^{M_{0}}\sum_{m=1}^{2}\rho _{m}\omega _{ml}C_{l}( \eta ,\xi ) }
\right\},  \label{fem0}
\end{equation}
where

\begin{equation}
C_{l}\left( \eta ,\xi \right) =\int \int_{A_{K}}\phi _{l}\left( \eta
^{\prime },\xi ^{\prime }\right) F\left( \eta ^{\prime },\xi ^{\prime },\eta
,\xi \right) d\eta ^{\prime }d\xi ^{\prime }.  \label{fem1}
\end{equation}
Evaluating Eq. (\ref{fem0}) at the $k$-th node $(\eta_k,\xi_k)$ and using
Eq. (\ref{coefficient1}), we get

\begin{equation}
\omega _{ik}=\exp \left\{ \mathrm{M}_{ik}+\sum_{l=1}^{M_{0}}\sum_{m=1}^{2}%
\rho _{m}\omega _{ml}C_{lk}\right\} -1,  \label{newton}
\end{equation}
with $C_{lk}\equiv C_{l}\left( \eta _{k},\xi _{k}\right) $ and M$_{ik}\equiv
$M$_{i}\left( \eta _{k},\xi _{k}\right) .$ Thus, we have a system of $2M_{0}$
non-linear algebraic equations which can be solved by any of the standard
methods, for example the Newton's method, which is our method of choice.

\subsection{Choice of the base functions $\phi_{l}\left( \eta ,\xi \right)$}

\label{sec.NumMet-Coord} To construct the base functions, it is necessary to
use a coordinate system defined in the element's domain. In our method, we
have used the area coordinates ($L_i$) defined as follows

\begin{equation}
L_{i}=\frac{a_{i}+b_{i}\eta +c_{i}\xi }{2\Delta }\hspace{0.2cm}\mathrm{with}%
\hspace{0.2cm}i=1,2,3,
\end{equation}
where $2\Delta$ is the area of the triangular element and

\begin{eqnarray}
a_{i} &=&\eta _{j}\xi _{k}-\eta _{k}\xi _{j},  \nonumber \\
b_{i} &=&\xi _{j}-\xi _{k}, \\
c_{i} &=&\eta _{k}-\eta _{j},  \nonumber
\end{eqnarray}
with cyclic rotation of indexes, where $j,k=1,2,3$ but $i\neq j\neq k$. The
set of points $\left\{ \left( \xi_{i},\eta _{i}\right) ,i=1,2,3\right\}$ are
the coordinates of the triangle corners. The relation between the
coordinates $\left( \xi ,\eta \right)$ and the triangular coordinates $%
\{L_i, i=1,2,3\}$ is given by

\begin{eqnarray}
\eta &=&L_{1}\eta _{1}+L_{2}\eta _{2}+L_{3}\eta _{3},  \nonumber \\
\xi &=&L_{1}\xi _{1}+L_{2}\xi _{2}+L_{3}\xi _{3}, \\
1 &=&L_{1}+L_{2}+L_{3}.  \nonumber
\end{eqnarray}

The number of nodes is equal to the number of base elements. A quadratic
base was used to solve Eq. (\ref{numerical1}) and therefore $L_{0}=6$. For
the corner nodes we have

\begin{eqnarray}
\phi _{1} &=&\left( 2L_{1}-1\right) L_{1}, \mathrm{\ etc.,}
\end{eqnarray}
and for the middle-side nodes

\begin{eqnarray}
\phi _{4} &=&4L_{1}L_{2}, \mathrm{\ etc.}
\end{eqnarray}
In this coordinate system, Eq.(\ref{fem1}) becomes

\begin{eqnarray}
C_{l}\left( \eta ,\xi \right) & = & 4 \Delta \int^{1}_{0}
\int^{1-L_2}_{0}\phi _{l}\left( L_{1},L_{2}\right)  \nonumber \\
&& \times F\left( L_{1},L_{2},\eta ,\xi \right) dL_{1}dL_{2}.  \label{fem2}
\end{eqnarray}


\newpage

\newpage

\end{document}